\documentclass[aps,notitlepage,twocolumn,footinbib,superscriptaddress,floatfix]{revtex4-1} 

\usepackage{CJK}
\usepackage{amsmath}
\usepackage{amsfonts}
\usepackage{bm}
\usepackage{soul}
\usepackage{comment}
\usepackage[caption=false,subrefformat=parens,labelformat=parens]{subfig}
\usepackage{color}
\usepackage[table,dvipsnames]{xcolor}
\usepackage[colorlinks,linktocpage,bookmarks=false]{hyperref}

\usepackage[normalem]{ulem} 
\definecolor{kspink}{RGB}{200,0,200}

\newcommand\myshade{85}
\definecolor{myrulecolor}{RGB}{150,20,0}
\colorlet{mylinkcolor}{violet}
\colorlet{mycitecolor}{YellowOrange}
\colorlet{myurlcolor}{Aquamarine}
\hypersetup{
	linkcolor  = myurlcolor!\myshade!black,
	citecolor  = mycitecolor!\myshade!black,
	urlcolor   =  myrulecolor!\myshade!black,
}

\usepackage{todonotes}

\usepackage{graphicx}
\graphicspath{{FIG/}}
\usepackage{multirow}
\usepackage{braket} 
\usepackage{booktabs} 

\AtBeginDocument{
	\heavyrulewidth=.08em
	\lightrulewidth=.05em
	\cmidrulewidth=.03em
	\belowrulesep=.65ex
	\belowbottomsep=0pt
	\aboverulesep=.4ex
	\abovetopsep=0pt
	\cmidrulesep=\doublerulesep
	\cmidrulekern=.5em
	\defaultaddspace=.5em
}

\newcommand{\figref}[1]{Fig.\,\ref{#1}}
\newcommand{\figsref}[1]{Figs.\,\ref{#1}}
\newcommand{\sfigref}[2]{Fig.\,\hyperref[#1]{\ref{#1}#2}}

\newcommand{\beq}{\begin{equation}}
\newcommand{\eeq}{\end{equation}}
\newcommand{\bea}{\begin{eqnarray}}
\newcommand{\eea}{\end{eqnarray}}

\newcommand{\ads}{$H_2\times S^1$}

\renewcommand\[{\begin{equation}}
\renewcommand\]{\end{equation}}

\newcommand{\treeon}{treeon}

\newcommand{\prlsection}[1]{\noindent\textit{\textbf{#1: }}}
\makeindex

\newcommand{\supp}{\textit{Supplementary Materials}\,}

\newcommand{\thistitle}{Y-cube model and fractal structure of subdimensional particles on hyperbolic lattices}

\begin{document} 
	\begin{CJK*}{UTF8}{gbsn} 
        \title{\thistitle}

		\author{Han Yan (闫寒)}
 	\email{hy41@rice.edu}
	\affiliation{Department of Physics \& Astronomy, Rice University, Houston, Texas 77005, USA}
  
		\author{Kevin Slagle} 
\affiliation{Department of Electrical and Computer Engineering, Rice University, Houston, Texas 77005 USA}
\affiliation{Department of Physics, California Institute of Technology, Pasadena, California 91125, USA}
\affiliation{Institute for Quantum Information and Matter and Walter Burke Institute for Theoretical Physics, California Institute of Technology, Pasadena, California 91125, USA}
		\author{Andriy H. Nevidomskyy} 
		\affiliation{Department of Physics \& Astronomy, Rice University, Houston, Texas 77005, USA}
 
		\date{\today}
\begin{abstract}
Unlike ordinary topological quantum phases, fracton orders are intimately dependent on the underlying lattice geometry.
In this work, we study a generalization of the X-cube model, dubbed the Y-cube model, on lattices embedded in  \ads\ space,
i.e., a stack of hyperbolic planes.
The name `Y-cube' comes from the Y-shape of the analog of the X-cube's X-shaped vertex operator.
We demonstrate that for certain hyperbolic lattice tesselations, the Y-cube model hosts a new kind of subdimensional particle, \textit{treeons}, which can only move on a fractal-shaped subset of the lattice. 
Such an excitation only appears on hyperbolic geometries;
on flat spaces treeons becomes either a lineon or a planeon.
Intriguingly, we find that for certain hyperbolic tesselations, a  fracton can be created by   a membrane operator (as in the X-cube model) \emph{or}   by a fractal-shaped operator within the hyperbolic plane.
\end{abstract}
\maketitle
\end{CJK*}


\prlsection{Introduction}
Fracton orders \cite{Pretko-review2020,NandkishoreReview,BravyiFracton,Haah2011,Vijay2015,Vijay2016,Chamon2005} are examples of highly entangled gapped phases of matter that lie beyond the Landau--Ginzburg paradigm in that no symmetry is broken, yet they are also distinct from the more familiar topological phases of matter in that they do not possess a universal long-wavelength description in terms of a topological quantum field theory (TQFT) \cite{Seiberg3,UVIRMixing,SlagleFQFT,foliatedDuality,ChamonQFTTypeII,Slagle-Kim2017,defectNetworks,WenCellular,JuvenCellular,stringMembraneNet,fractonHigherForm}.
Rather, fracton orders have a nontrivial ground state degeneracy that depends not only on the topology but also on the system size \cite{BravyiFracton,Haah2011,Vijay2015,Vijay2016} and lattice geometry \cite{ShirleyFoliation,Slagle-Kim2017}.
Furthermore, fracton orders support excitations whose mobility is restricted when we do not allow any additional excitations to be created:
  fractons are immobile; lineons can only move along a one-dimensional line, and
  planeons can only move within a two-dimensional plane~\cite{PaiHermeleFusion,extendedFracton}. 

Much attention has been devoted to exactly solvable models that host fracton order in flat space, such as the X-cube model formulated on the cubic lattice \cite{Vijay2016}. By comparison, relatively little is known about the behavior of such models in curved spaces  \cite{HanFractonHolography,HanFractonHolography2,HanFractonHolography3,ShuHengComplexityGraphs,ShuHengLineonGraphs,RadicevicSystematic} (see Refs.~\cite{Slagle2019,ProhazkaFractonsCurved,JensenFractonsCurved,GromovCurved} for works that study gapless fracton models \cite{Pretko2017,SeibergU1,ElasticityDuality,gaplessChernSimons} 
on curved spaces ). 
The fundamental motivation for introducing curvature is to investigate how it affects the properties of the fracton order, in much the same way as placing a TQFT on a manifold with a different genus teaches us about the topological nature of the ground state degeneracy.
For example, it has been previously noted that curvature can lead to 
a robust ground state degeneracy of X-cube model even on manifolds that are topologically trivial \cite{Slagle-Kim2017};
and
curvature can grant the subdimensional particles additional mobility \cite{SlagleLattices,SlagleCurvedU1}.
Another practical motivation for introducing curvature is to search for better error correcting fracton codes.
In particular, codes on hyperbolic spaces 
can have favorable quantum error correcting properties \cite{BalancedProduct}.
It is thus interesting to examine this aspect for fracton order \cite{ZhenghanHaah,TianHaah}.

\begin{figure}
	\centering
 \subfloat[\label{Fig_3d_54}(5,4) tessellation]{\includegraphics[width=0.8\columnwidth]{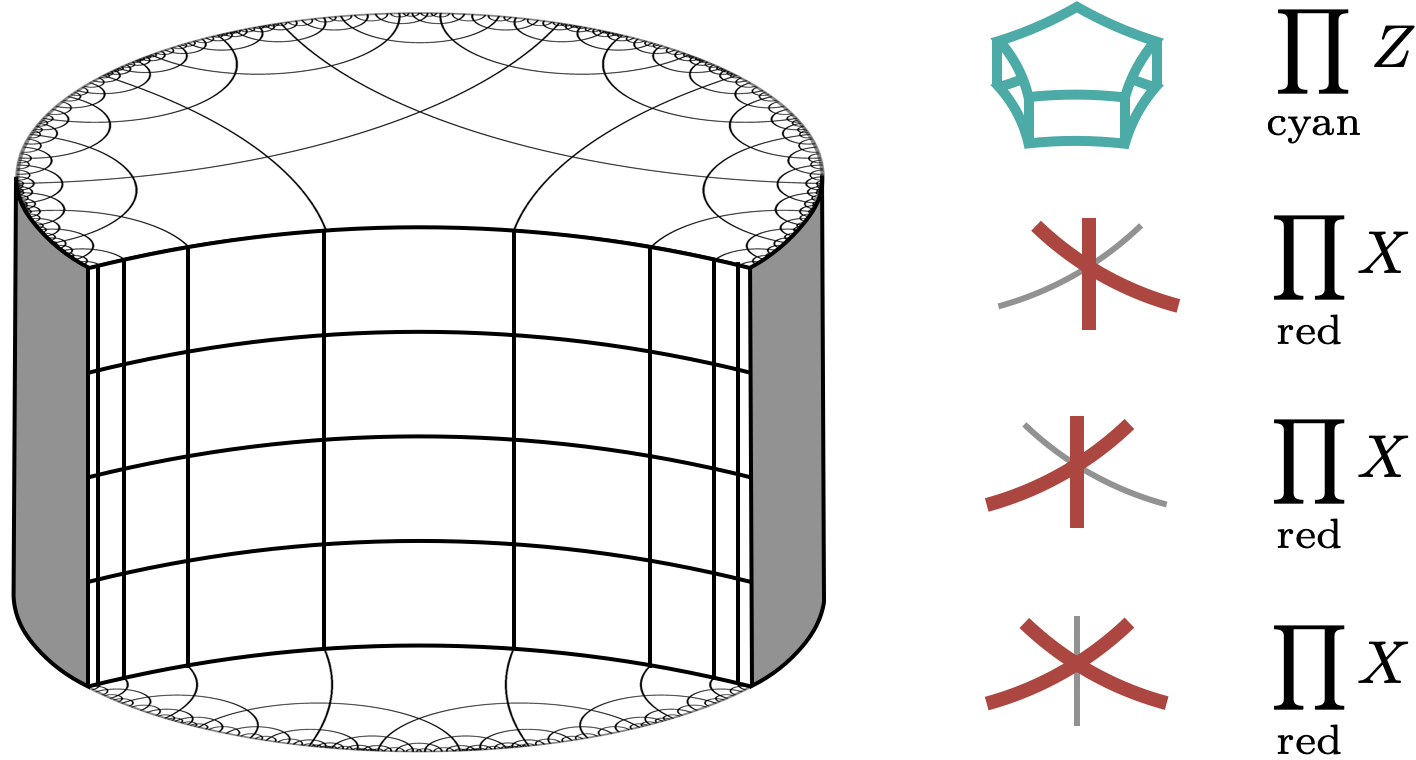}}	\\
  \subfloat[\label{Fig_3d_46}(4,6) tessellation]{\includegraphics[width=0.8\columnwidth]{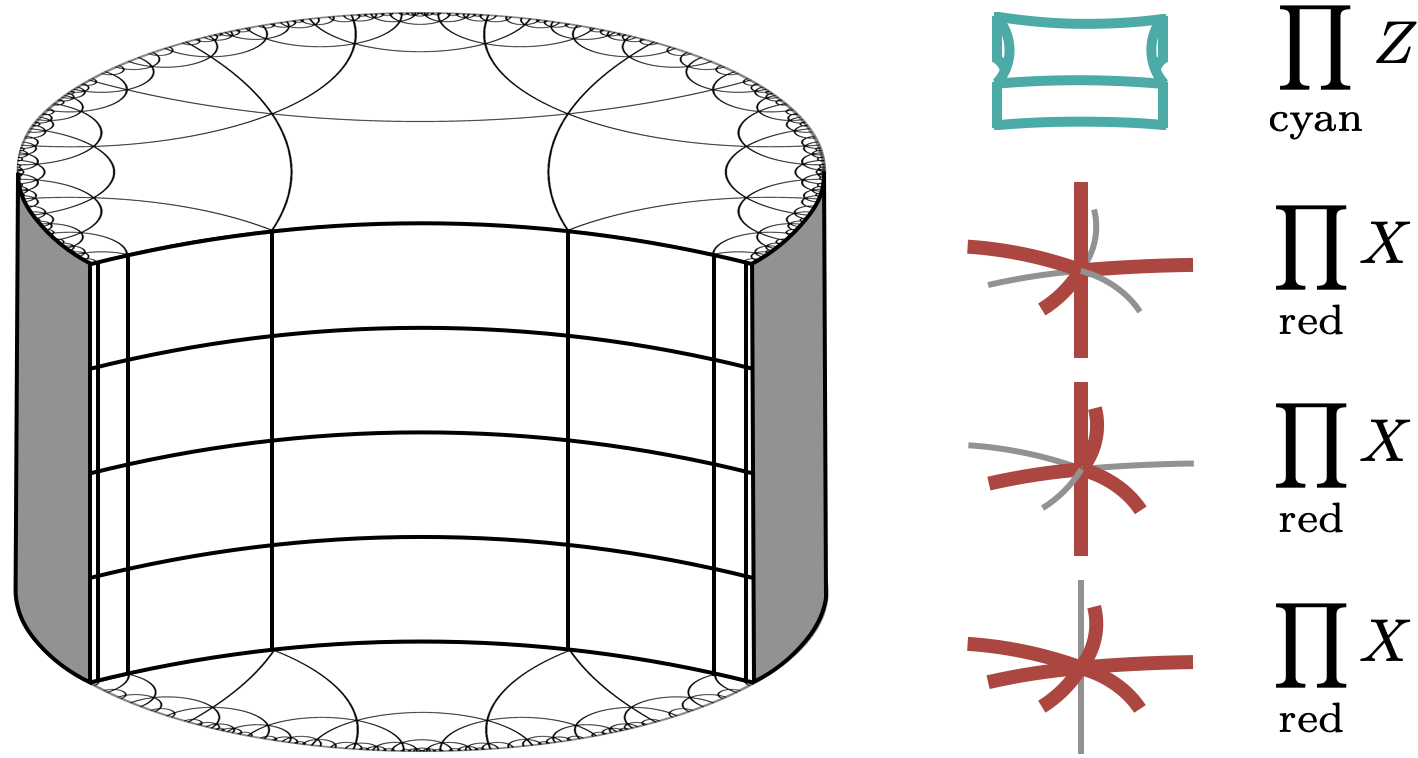}}	
	\caption{\label{Fig_hyperbolic_X_cube}Examples of tessellations of the \ads manifold using the Poincar\'e disk representation:
all polygons on the disk have identical area, but look smaller when drawn farther from the center. 
(a) Hyperbolic   tessellation with $(p,q)=(5,4)$ (left) and Hamiltonian terms (right).
(b) Hyperbolic   tessellation  with $(p,q)=(4,6)$  (left) and Hamiltonian terms (right).
} 
\end{figure}

In this Letter, we investigate a generalization of the X-cube model to 
the hyperbolic space $H_2\times S^1$, which can be visualized as a stack of hyperbolic planes ($H^2$) with the top and bottom layers identified. Unlike the flat space, which only permits a small number of different lattices viewed as tessellations by regular polygons/polyhedra (i.e. the familiar square, triangular and hexagonal lattices in two dimensions),
the number of distinct tessellations is infinite in hyperbolic spaces.
Regular two-dimensional hyperbolic plane tessellations are enumerated by a pair of integers $(p,q)$ satisfying $\frac{1}{p} + \frac{1}{q} < 1/2$,
which is called the Schl\"{a}fli symbol.
These tessellations consist of
of $p$-gonal regular polygons, with $q$ polygons meeting at each vertex. 
%
Two examples of such tessellations are shown in \figref{Fig_hyperbolic_X_cube}.

We find that the generalized X-cube model on
this hyperbolic geometry depends sensitively on the tessellation. The simplest is the case of $q\!=\!4$, where each vertex is locally isomorphic to that of a cubic lattice, allowing for the standard definition of the vertex  operators as products of four Pauli $X$ in each of the three locally orthogonal intersecting planes. The resulting $(p,4)$ model has one-dimensional (1D) particles, lineons, which propagate along the geodesics of the $H_2$ plane (as opposed to straight lines in the flat space), but otherwise are very similar to the X-cube lineons. There are nuances with the operators necessary to create individual fractons however, as we shall see below. 

\begin{figure}
	\centering
\subfloat[\label{Fig_46tess_lineon_pair}]{\includegraphics[width=0.33 \columnwidth]{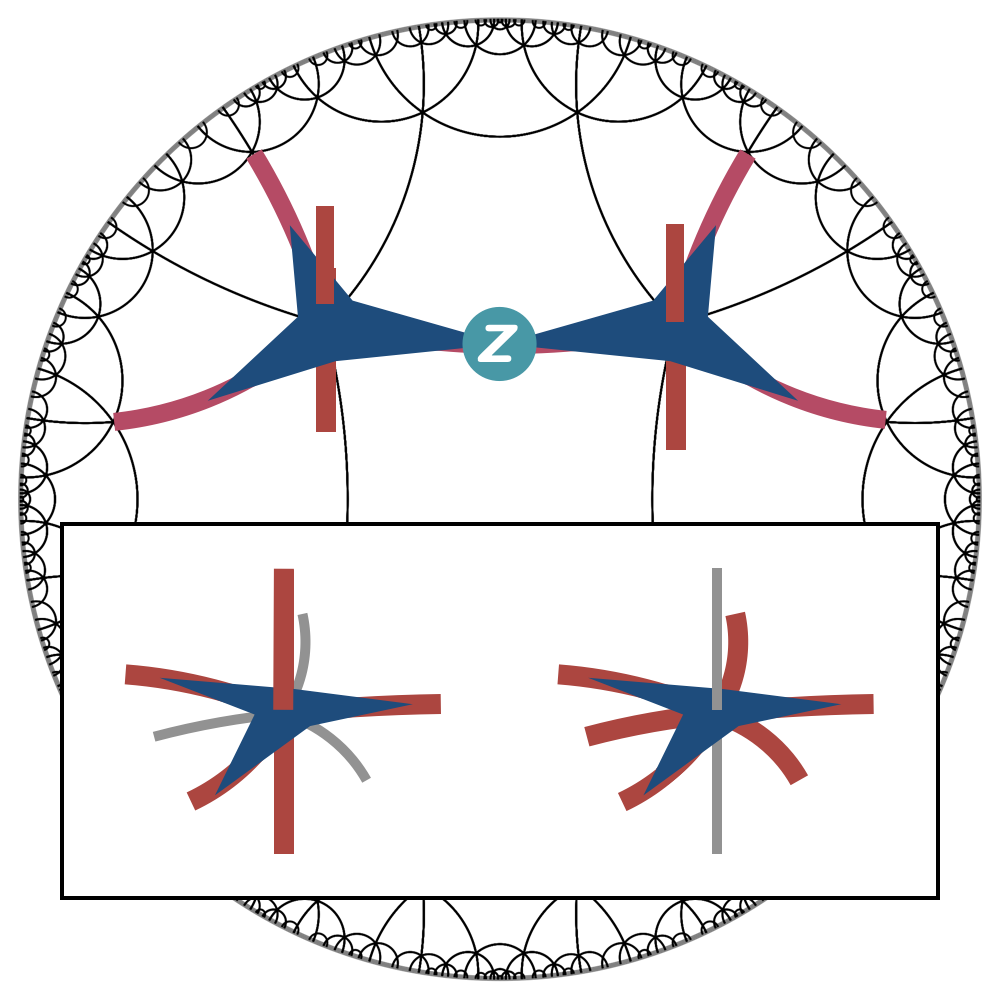}}	
 \subfloat[\label{Fig_46tess_lineon_move}]{\includegraphics[width=0.33 \columnwidth]{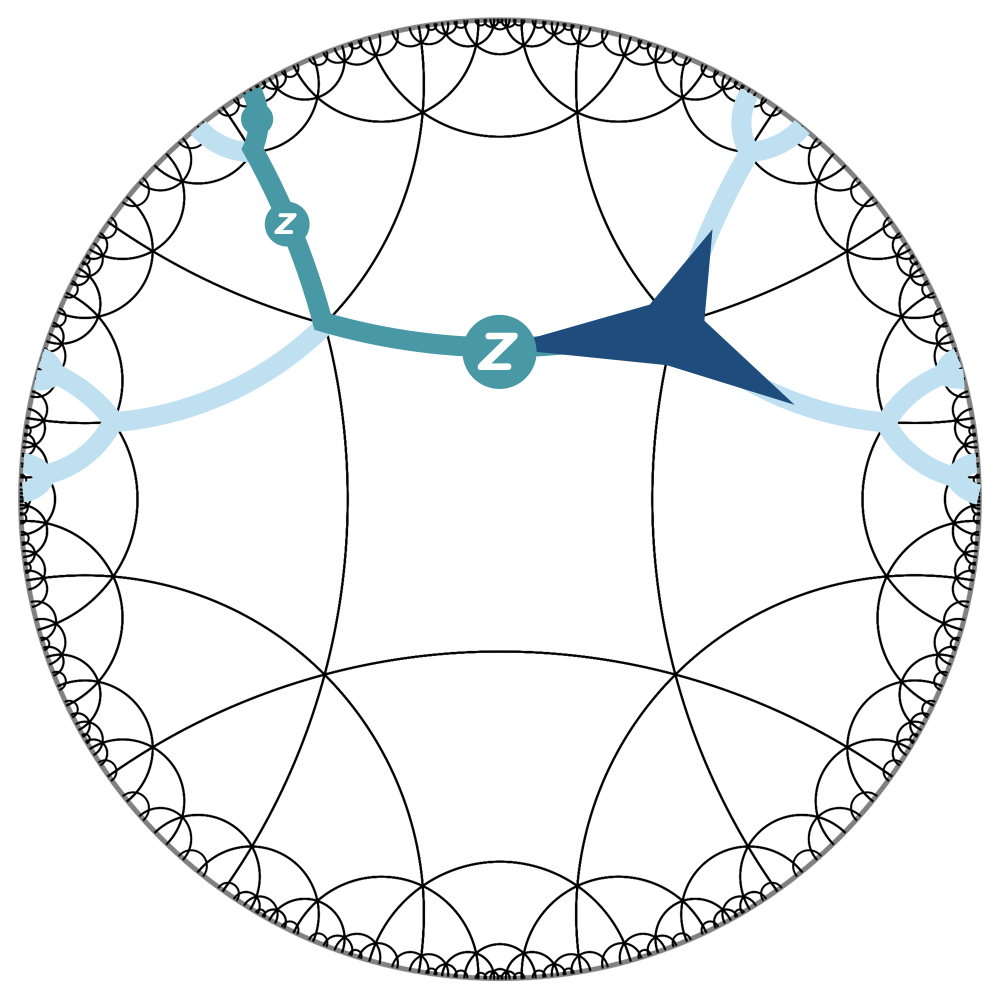}}
\subfloat[\label{Fig_46tess_lineon_logical}]{\includegraphics[width=0.33 \columnwidth]{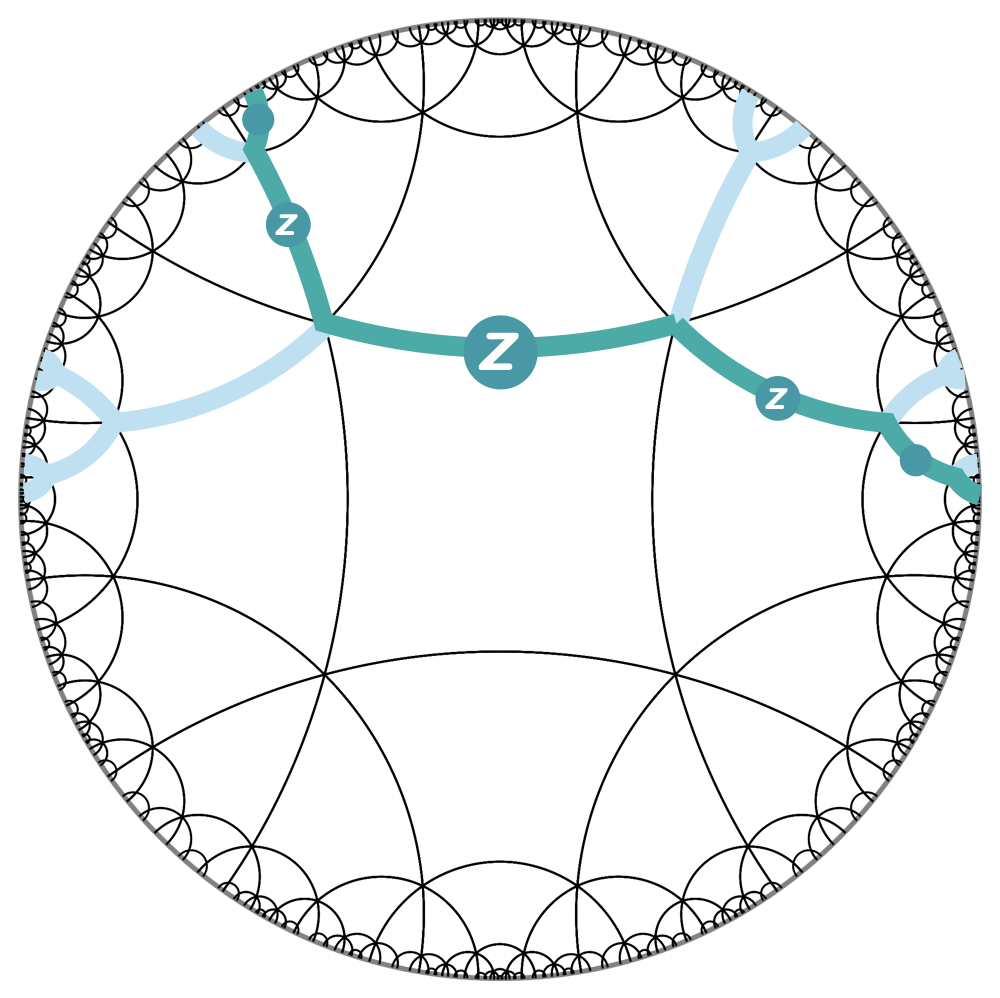}}	
\caption{\label{Fig_46tess_z}
 Treeon operators for the $(6,4)$ tessellation:
(a) A $Z$ operator (teal) on an in-plane edge creates two treeons (blue stars).
The inset shows the two excited terms in the Hamiltonian for a single treeon.
(b) Applying a string of $Z$ operators (colored  teal)  within  the fractal tree (colored light teal) creates a single \treeon.
The \treeon\ can only move within the fractal tree.
(c) An infinite string of $Z$ operators within the fractal tree is a logical operator.
} 
\end{figure}

The most intriguing findings are for tessellations of $q>4$.
For even $q>4$, we find  new models, which we dub \textit{Y-cube} (because of the Y shape of the in-plane part of two of the vertex operators in the simplest $q=6$ case illustrated in \figref{Fig_3d_46}). Unlike the X-cube model, the Y-cube model with $q>4$ does not possess lineons; instead the lineons are replaced with a new kind of quasiparticles, \textit{treeons}, that can only propagate on a fractal tree as shown in \figref{Fig_46tess_lineon_move}.
Moreover, a pair or ``dipole'' of neighboring fractons remains immobile within the $H_2$ plane, in contrast to X-cube model where fracton dipoles forms a planeon. \\


\prlsection{X-cube and Y-cube models in \ads}
The generalized X-cube models are constructed as shown in \figref{Fig_hyperbolic_X_cube}.
The Hamiltonian consists of two types of terms: the vertex and the prism (generalization of the cube) terms.
For $(p,q=4)$ tessellations, the model is the natural generalization of the X-cube model \cite{ShirleyFoliation,SlagleLattices}:
  the vertex terms are identical to those in the X-cube model, while the ``cube" terms become products of $Z$ operators over the edges of the $p$-gonal prisms, as shown in \figref{Fig_3d_54}.

In the general case of $(p,q)$ tessellations (with even $q$), 
we keep the $p$-gonal prism $Z$-operators in the Hamiltonian. 
There are in general two kinds of vertex terms:
(1) a product of  $X$ operators on the two neighboring out-of-plane  edges   and $q/2$ nonadjacent in-plane edges neighboring the vertex; and
(2) a product of   $X$ operators on the $q$ in-plane edges  neighboring the vertex.
See \figref{Fig_3d_46} for a $q=6$ example.
Each vertex term has an even number of X operators that overlap with a prism Z operators, making the model a stabilizer code 
Hamiltonian.
%
We name this model the \textit{Y-cube model}, alluding to the ``Y'' shape of 
the in-plane vertex terms when $q=6$ [\figref{Fig_3d_46}].

\begin{figure}
	\centering
\subfloat[\label{Fig_54tess_4fracton}]{\includegraphics[width=0.33 \columnwidth]{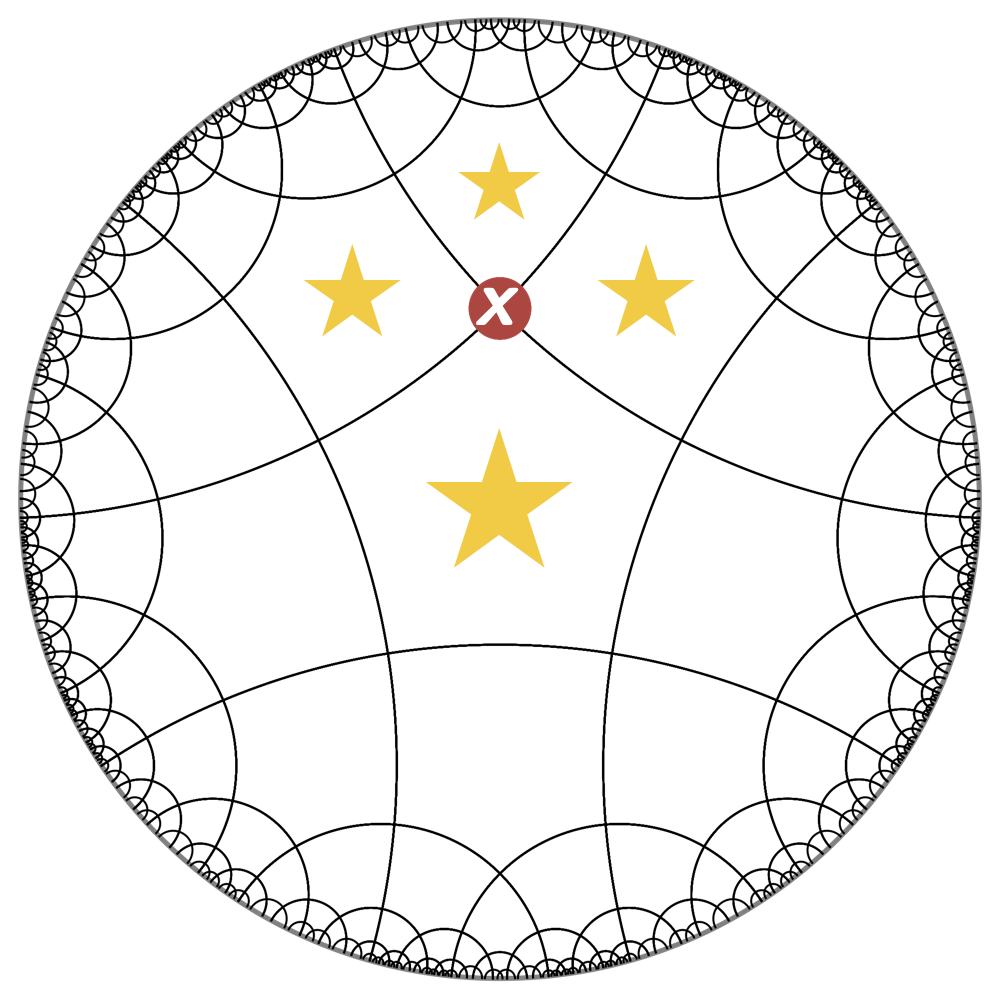}}	
 \subfloat[\label{Fig_54tess_2fracton}]{\includegraphics[width=0.33 \columnwidth]{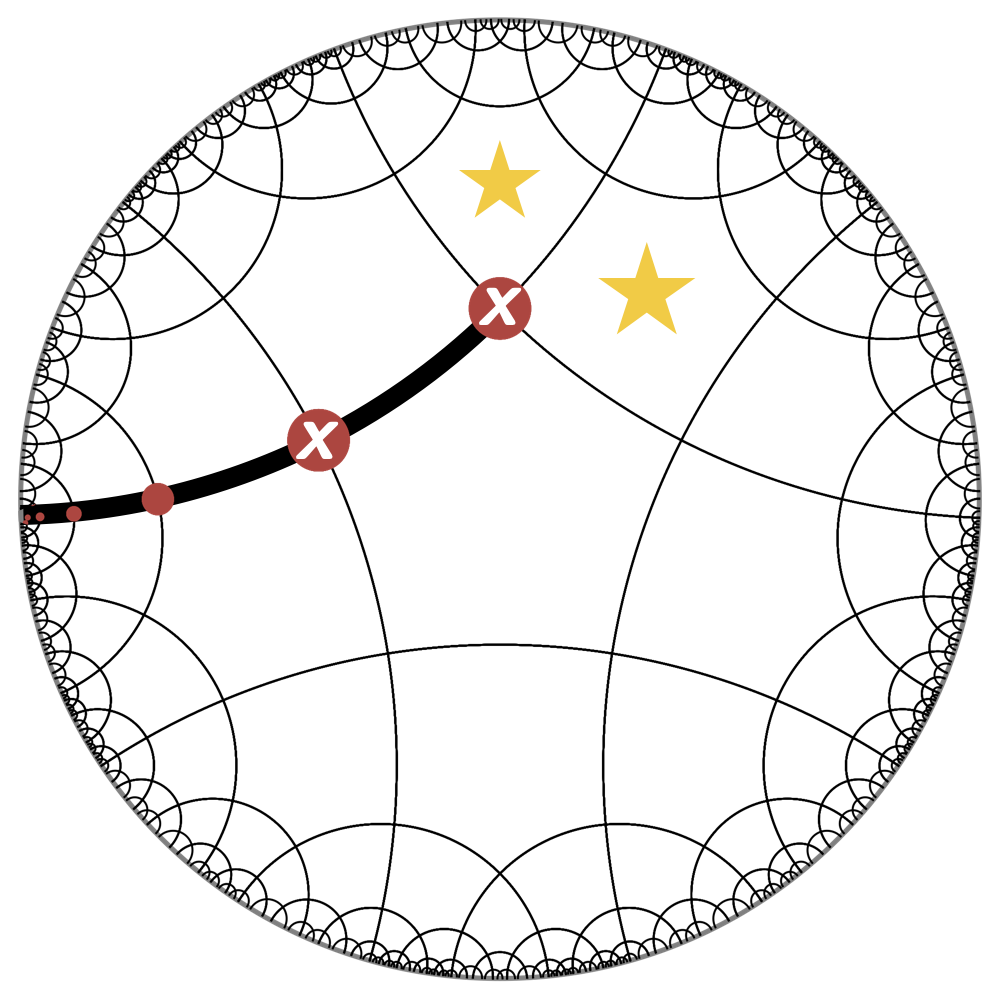}} 	
 \subfloat[\label{Fig_54tess_1fracton}]{\includegraphics[width=0.33 \columnwidth]{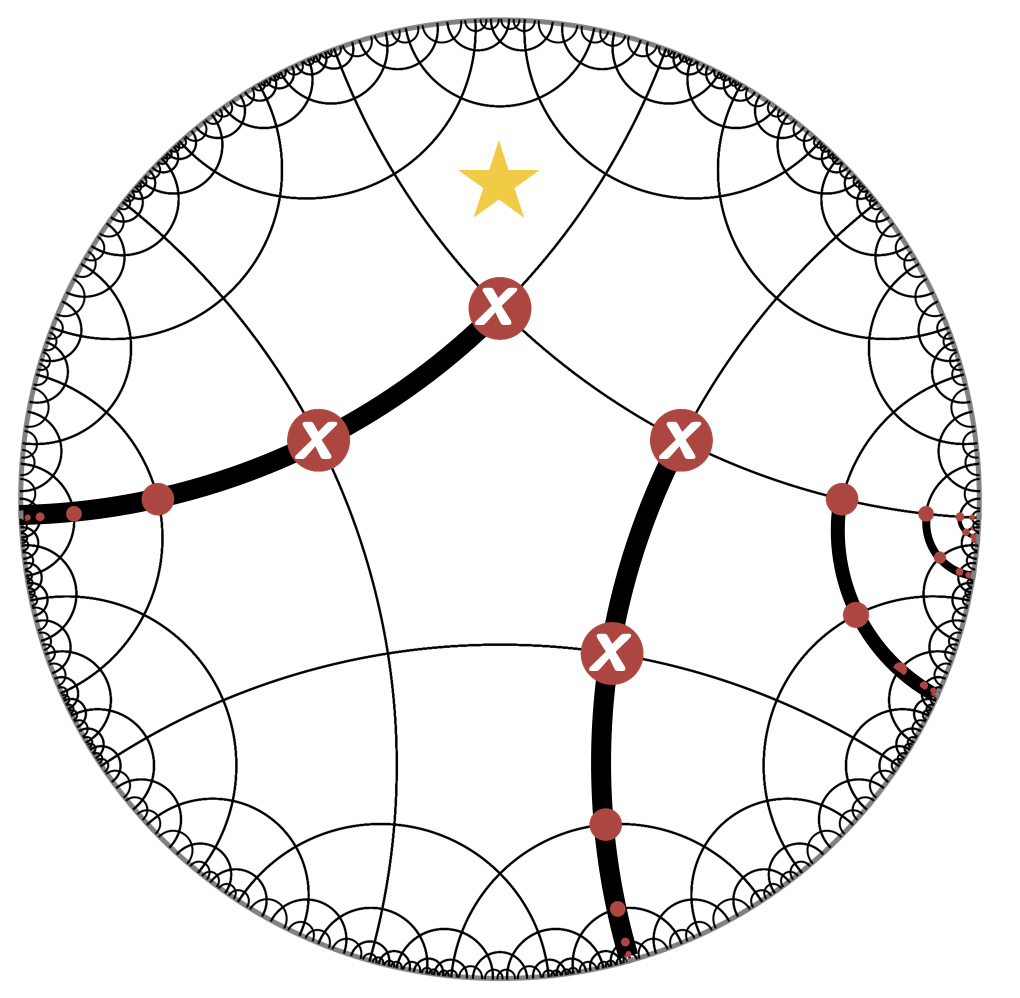}}	
	\caption{	\label{Fig_54tess_x}
 Fracton operators for the $(5,4)$ tessellation:
 (a) An $X$ operator  (red dot) on an out-of-plane edge (perpendicular to the shown hyperbolic plane) creates four fractons (yellow stars).
 (b) A truncated geodesic of $X$ operators creates two fractons, which can move along the geodesic.
 (c) $X$ operators on a 
  stack
 of truncated geodesics create a single fracton.
	} 
\end{figure}

Before we move on to describe the new features of the hyperbolic X/Y-cube models, 
we briefly note that all of them share some common properties due to the flat $S^1$ dimension. 
Acting with an $X$ operator on an in-plane edge will create four fracton excitations in the four prisms  neighboring the edge.
A pair of fractons displaced out of the plane is mobile 
  within the hyperbolic plane
  (via $X$ operators acting on the in-plane edges).
A pair of fractons displaced in-plane (see e.g. \figref{Fig_54tess_2fracton}) is a lineon that can move in the out of plane direction.
A pair of fractons displaced out of the plane is mobile 
within the hyperbolic plane
(via $X$ operators acting on the in-plane edges).
Acting with a $Z$ operator on an out-of-plane edge will create two lineons on the two vertices at the ends of the edge. These lineons are free to move in the out-of-plane direction.
These excitations are similar to excitations in the cubic lattice X-cube model. 
However the excitations (fractons, lineons, and the new treeons) created otherwise---via $X$ operators acting on a out-of-plane edges or $Z$ operators on in-plane edges---have new physics that depends on the tessellation, as we discuss in detail below. \\

\prlsection{Fractons in $(5,4)$ X-cube   model} 
Let us first consider the model on the hyperbolic lattice  with  $(p,q) = (5,4)$ [\figref{Fig_3d_54}], 
whose physics  generalizes straight-forwardly to all \mbox{$(\text{odd }p\geq5,q = 4)$} tessellations.

We first examine the effects of an $X$ operator acting on an out-of-plane edge.
It creates four fractons (a quadrupole) on the  four  neighboring prisms [\figref{Fig_54tess_4fracton}].
By consecutively applying $X$ operator on the out-of-plane edges attached to the same geodesic of the $H_2$ plane, 
a pair (or dipole) of the fractons can be moved away. 
Extending one side of the string to the infinite boundary of the hyperbolic plane will leave a single pair of fractons in the bulk.
Equivalently, a truncated geodesic of $X$ operators creates a pair of fractons at its end [\figref{Fig_54tess_2fracton}].

Unlike in the X-cube model,
  a single fracton cannot be created in the bulk at the corner of a membrane operator.
This is because a membrane operator  creates fractons inside the membrane since each pentagon prism is surrounded by an odd number ($p=5$) of out-of-plane edges.
Instead, a single fracton can be created using a series of truncated geodesic strings of $X$ operators, as illustrated in \figref{Fig_54tess_1fracton}.
Each truncated geodesic of out-of-plane $X$ operators creates a pair of fractons. 
Thus, the first truncated geodesic operator creates a pair of fractons, and the others moves one fracton in the pair away to infinity.
\\

\begin{figure}
	\centering
\subfloat[\label{Fig_54tess_lineon_pair}]{\includegraphics[width=0.33 \columnwidth]{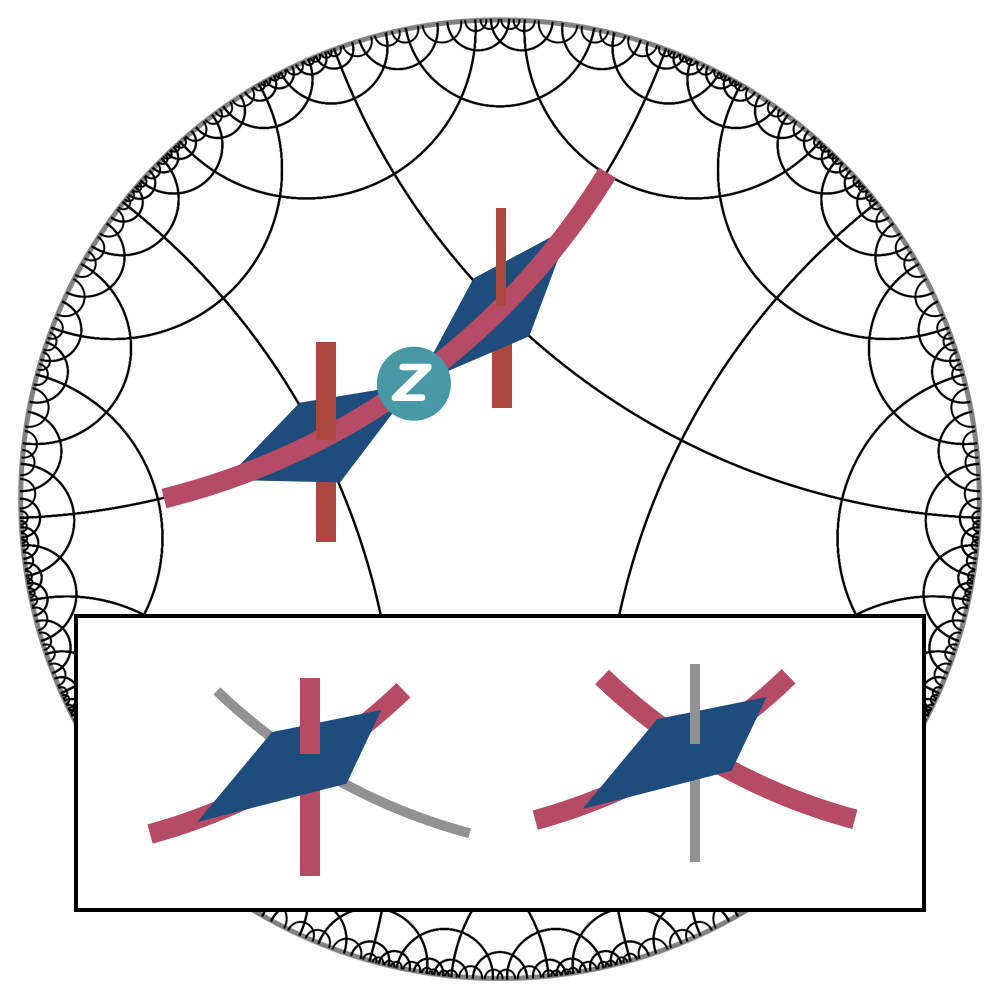}}	
 \subfloat[\label{Fig_54tess_lineon_move}]{\includegraphics[width=0.33 \columnwidth]{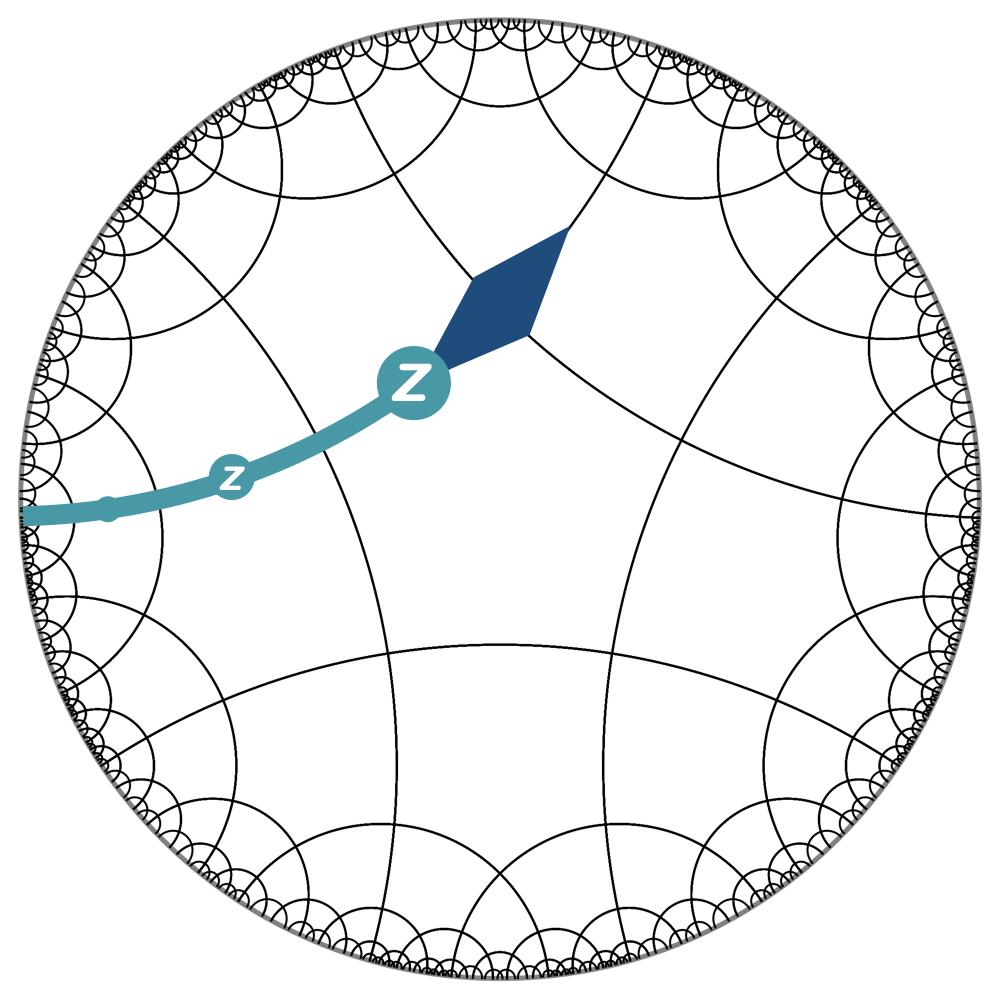}} 
 \subfloat[\label{Fig_54tess_lineon_line}]{\includegraphics[width=0.33 \columnwidth]{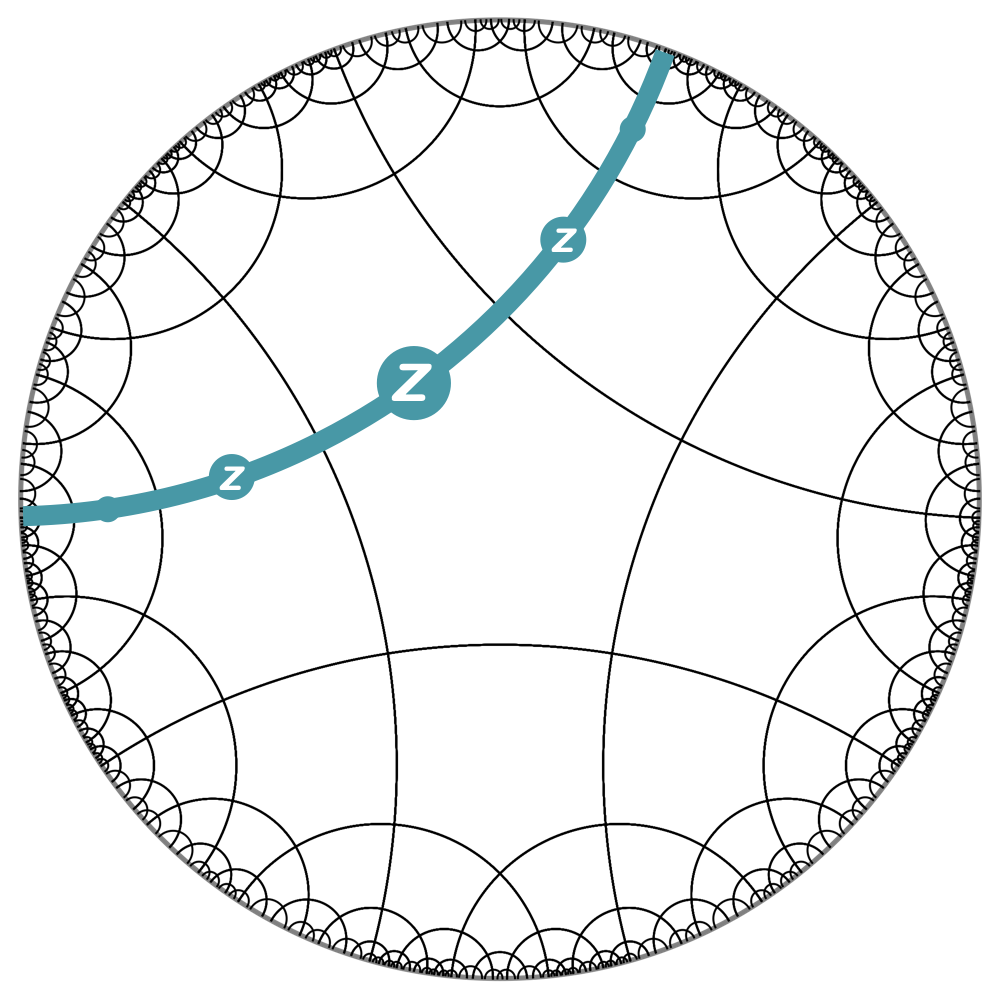}}	
	\caption{	\label{Fig_54tess_z}
 Lineon operators for the $(5,4)$ tessellation:
 (a) A $Z$ operator (teal) on an in-plane edge creates two lineons (blue diamonds).
The inset shows the two excited terms in the Hamiltonian for a single lineon.
(b) Applying a string of  $Z$ operators along a geodesic of in-plane edges creates a single lineon, which can move along the geodesic.
(c) The logical operator constructed by $Z$ operators, which can also be viewed as moving a lineon from one boundary to the other.
} 
\end{figure}

\prlsection{Lineons in $(5,4)$ X-cube   model}  
Next we examine the action of $Z$ operators on in-plane edges. 
When $q=4$, each vertex is locally identical to a vertex of the cubic lattice. 
Hence, a  $Z$ operator on an in-plane link creates two  lineons [\figref{Fig_54tess_lineon_pair}] that behave  similarly to lineons in the X-cube model on a cubic lattice. 
Each lineon is an excited state of two vertex operators, shown in the inset of  \figref{Fig_54tess_lineon_pair}.
Lineons are restricted to move on a $H_2$ geodesic, as shown in \figref{Fig_54tess_lineon_move}.
Under rough boundary conditions (which condense lineons) of the hyperbolic planes,
the product of 
$Z$ operators 
along a geodesic [\figref{Fig_54tess_lineon_line}]
becomes a logical operator that does not create any excitations.
\\
 
\begin{figure} 
	\centering
  \subfloat[\label{Fig_46tess_fractal_logical}]{\includegraphics[width=0.33 \columnwidth]{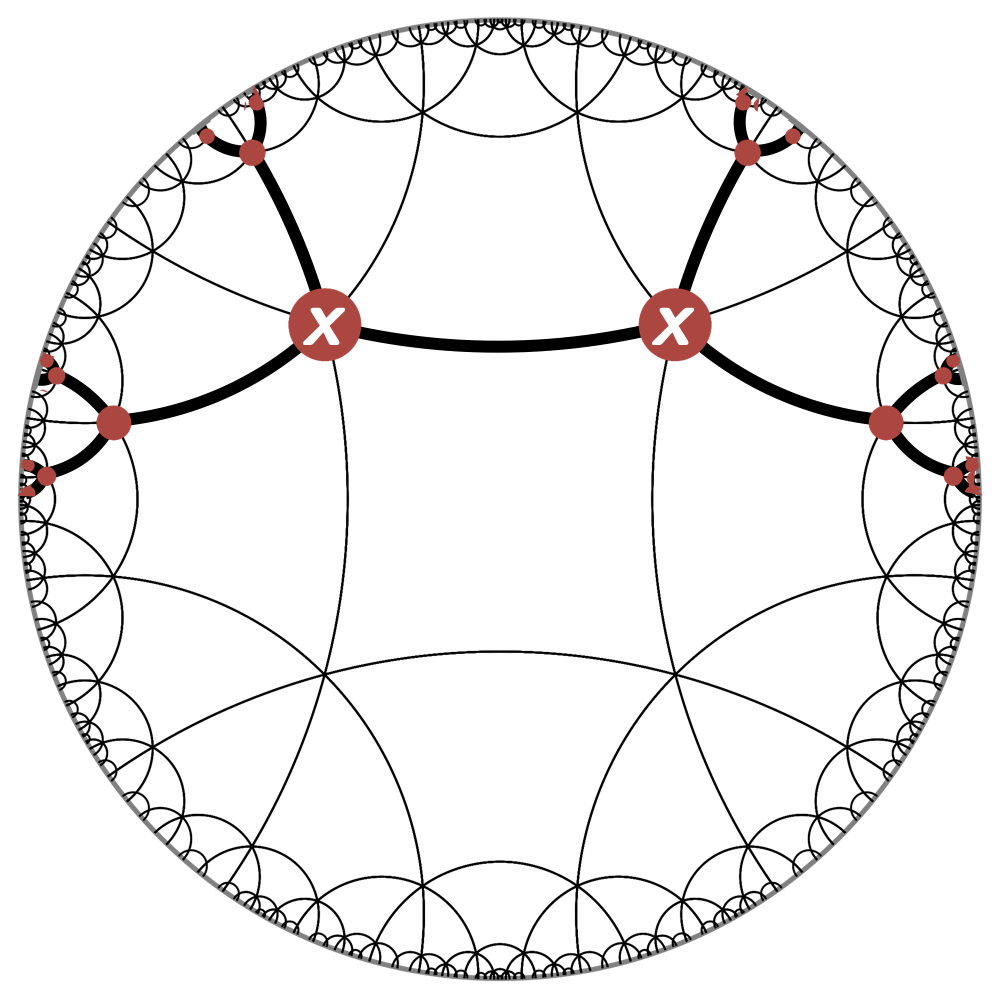}}	
 \subfloat[\label{Fig_46tess_fractal_2fracton}]{\includegraphics[width=0.33 \columnwidth]{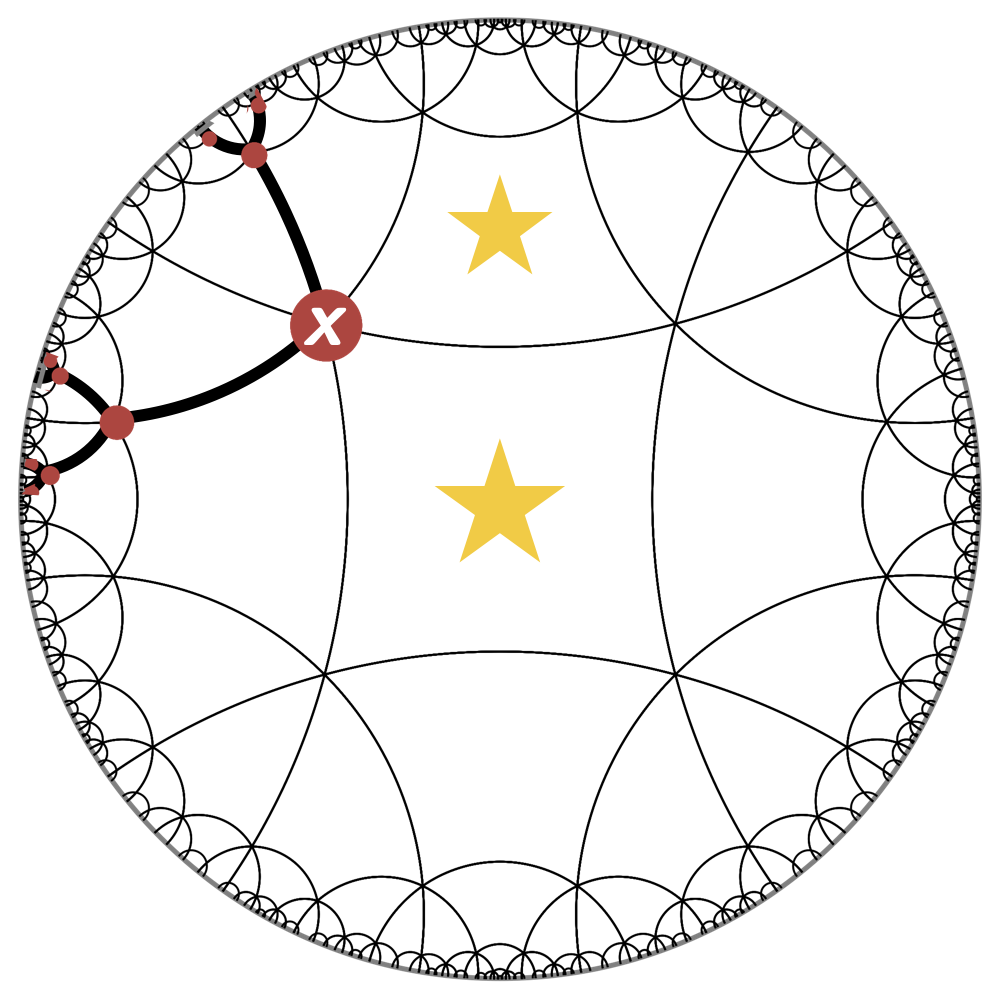}} 
\subfloat[\label{Fig_46tess_fractal_1fracton}]{\includegraphics[width=0.33 \columnwidth]{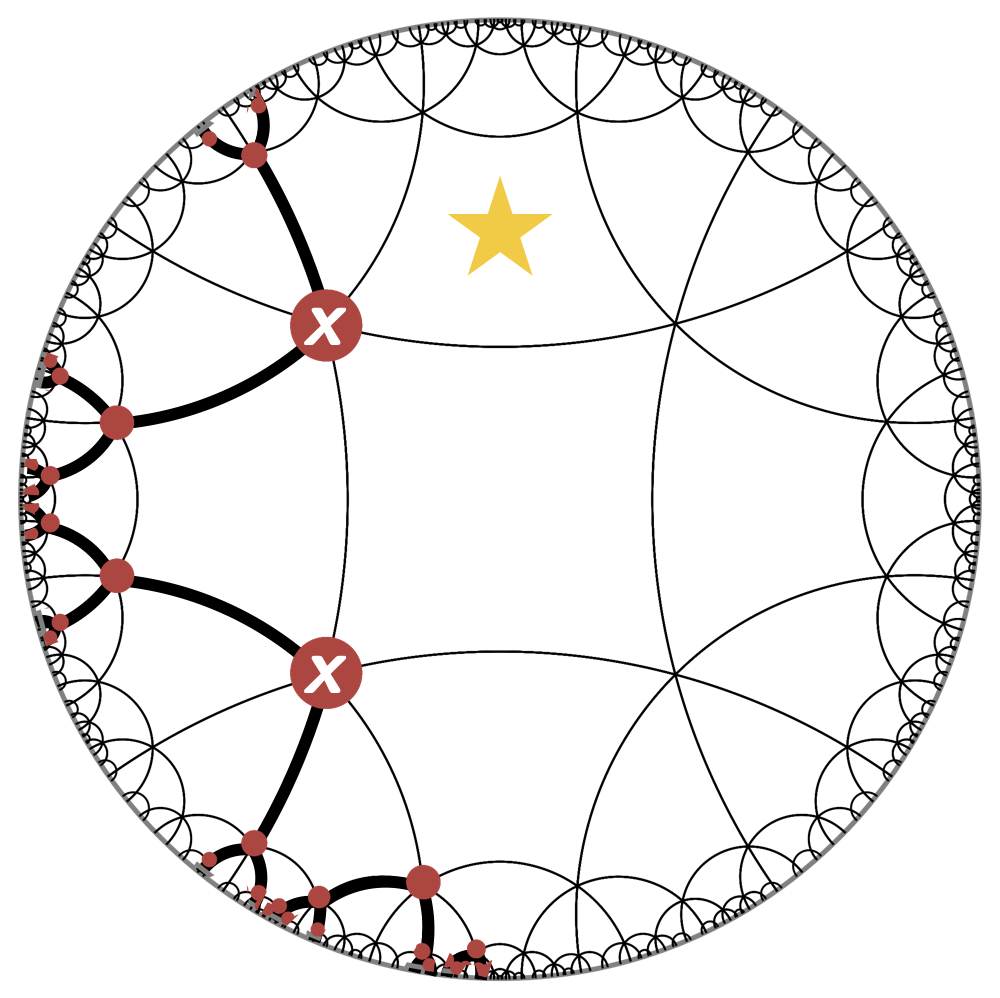}}	\\
\subfloat[\label{Fig_46tess_x_topo}]{\includegraphics[width=0.33 \columnwidth]{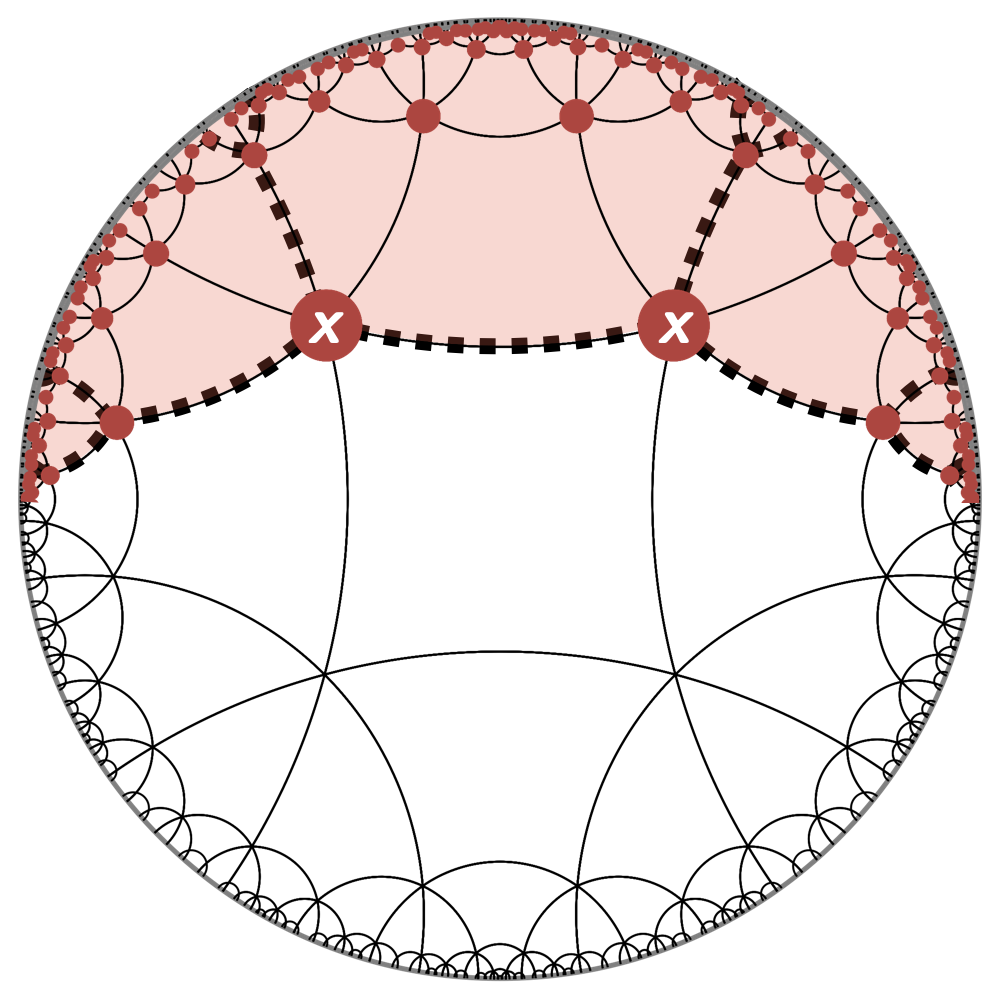}}
 \subfloat[\label{Fig_46tess_1fracton}]{\includegraphics[width=0.33 \columnwidth]{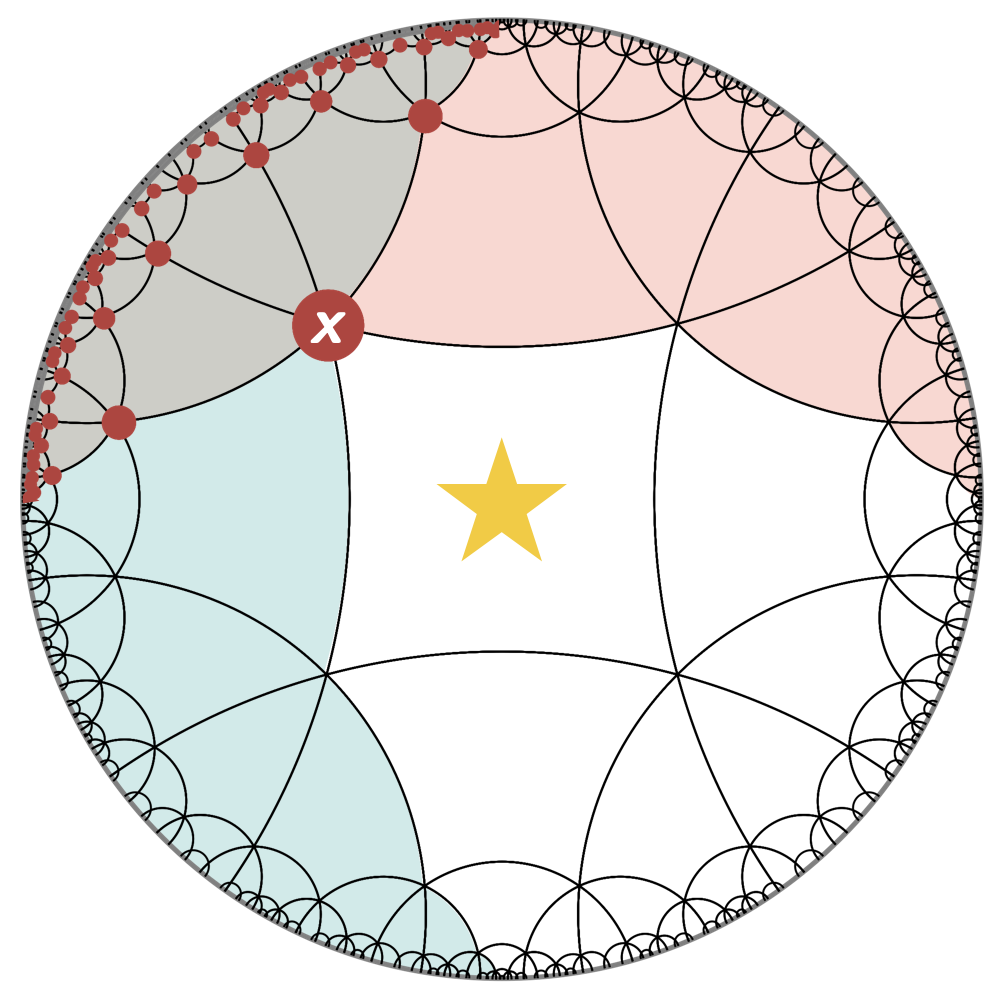}}
\caption{	\label{Fig_46tess_x}
 Fracton operators for the $(4,5)$ tessellation:
(a) A bulk logical  $X$ operator on a fractal tree (thick black),
  which is a product of $X$ operators (red dots) on out-of-plane edges neighboring one (out-of-plane) side of the fractal tree.
(b) A pruned fractal-tree of $X$ operators creates a pair of fractons, which is a lineon with mobility only in the out-of-plane direction.
(c) An infinite series of the pruned fractal-trees creates a single fracton.
(d) A bulk logical  $X$ operator on a fractal tree \textit{wedge} (colored red). It is the product of all $X$ operators (red dots) on the out-of-plane edges neighboring one side of the wedge.
The fractal tree is  drawn in thick, dashed line.
(e) A membrane of $X$ operators supported on the intersection of two wedges (red and teal) also creates a single fracton. 
} 
\end{figure}

\prlsection{Fractons in $(4,6)$ Y-cube   model} 
Next we consider the Y-cube  model  on  $(\text{even } p\ge4,q \ge 6) $ tessellations.
We find that $q \ge 6$ results in novel physics with a new kind of restricted particle mobility.
We focus on the representative example of $(4,6)$, with the Hamiltonian shown in \figref{Fig_3d_46}.
We first discuss the properties of fractons, then we consider the  treeons (which are analogs of lineons).

On the $(4,6)$ tessellation, there are two ways to create fractons, illustrated in \figref{Fig_46tess_x}. 
To understand the first way, it is helpful to first construct logical out-of-plane
$X$ operators that do not create any fractons in the bulk, shown in \figref{Fig_46tess_fractal_logical}.
The logical operator is a product of $X$ operators
on a fractal tree (also known as a Bruahat-Tits tree).
The fractal tree is constructed by choosing $q/2$ non-adjacent edges at a vertex and repeating this procedure at every vertex it extends to.
We shall refer to these logical operators as $T_X$.
The $T_X$ operators anti-commute with strings of $Z$ operators in the out-of-plane direction.
Here, we assume either an infinite hyperbolic plane or a finite plane with a boundary that condenses fracton dipoles \cite{BulmashBoundary,KarchBoundary}.

Now we can see how single fracton and fracton dipoles are created by the \textit{first type} of fractal tree operator:
an in-plane fracton dipole is created by pruning the fractal tree operator $T_X$ at a vertex in the bulk,
as shown in \figref{Fig_46tess_fractal_2fracton}.
Unlike the X-cube model on a cubic lattice or a $(p,q=4)$ tessellation, for even $q\geq6$ the
  in-plane fracton dipoles are lineons that can only move in the out-of-plane direction.
A single fracton can be created by aligning many pruned fractal trees along a series of adjacent vertices,
as shown in \figref{Fig_46tess_fractal_1fracton}.
Each pruned tree, creating a dipole, 
serves the purpose of moving a fracton closer to the boundary. 
%

Now we turn to a \textit{second type} of fracton-creation operator.
Recall that when $p$ is odd (such as $p=5$ in the earlier example in \figref{Fig_54tess_x}), a membrane of $X$ operators creates an extensive number of excitations
%
because each prism term overlaps with the membrane operate by an odd number ($p$) of out-of-plane edges. 
However when $p$ is even, $X$ membrane operators commute with the prism term in the bulk. 
%

The key to constructing such membrane operators is that its boundary should overlap  with the neighbouring prisms by two out-of-plane edges. 
Following this principle, the membrane 
boundary should follow this pattern:
we first find a $q/2$-degree fractal tree   we constructed earlier. 
Then, we select a region bounded by the branches of the tree that contains the entire tree [shaded region in \figref{Fig_46tess_x_topo}].
The membrane   operator is the product of all the $X$ operators on the out-of-plane edges attached to this region (on the top or bottom side of the $H^2$ plane).
We name this geometric shape the \textit{fractal-tree wedge}.


To construct a membrane operator that creates a single fracton,
  we can select two partially overlapping fractal-tree wedges.
The membrane operator supported on the overlap    creates a single fracton near the intersection of wedge boundaries, as shown in  \figref{Fig_46tess_1fracton}.
\\

\begin{table*}[t!]
\caption{
Properties of hyperbolic X/Y-cube models on different tessellations of \ads
\label{TABLE_hyperbolic_X}
} 
\begin{tabular}{c @{\hskip 10pt}  c @{\hskip 10pt} c @{\hskip 10pt} c @{\hskip 10pt} c  } 
	\toprule
	lattice and model 
 &  \begin{tabular}{@{}c@{}} 
	 in-plane \\
   lineon/\treeon 
	\end{tabular}
  & \begin{tabular}{@{}c@{}} 
	 in-plane \\
   fracton dipole 
	\end{tabular}   
 & \begin{tabular}{@{}c@{}} 
	 in-plane \\
     $X$ logical op.
	\end{tabular}  
   & \begin{tabular}{@{}c@{}} 
	 in-plane \\
     fracton creation op.
	\end{tabular}     
   \\  \midrule\midrule
	$(p \text{ odd},q = 4)$ 
    X-cube   
    &     lineon, \figref{Fig_54tess_lineon_move}
    &     1D mobility, \figref{Fig_54tess_2fracton}
    &
	geodesic  
    &   
	truncated geodesics,   \figref{Fig_54tess_2fracton} \\
	\midrule
	$(p \text{ even} , q = 4)$   
    X-cube
    &        lineon  
    &     1D mobility 
    & \begin{tabular}{@{}c@{}} 
	 geodesic, \\
     geodesic wedge, \figref{Fig_64tess_x_topo}
	\end{tabular}  
    &   \begin{tabular}{@{}c@{}} 
	 truncated geodesics, \\
     wedge corner, \figref{Fig_64tess_fracton}
	\end{tabular} 
    \\ \midrule
	$(p \text{ odd},q   \ge 6)$ 
    Y-cube    
    &    treeon     
    &   no  mobility  
    & 
	fractal tree 
    &    
	pruned fractal trees
	\\  \midrule
	$(p \text{ even}, q   \ge 6)$   
    Y-cube    
    &      treeon, \figref{Fig_46tess_lineon_move}   
    &      no  mobility, \figref{Fig_46tess_fractal_2fracton}
    & \begin{tabular}{@{}c@{}} 
	fractal tree,  \figref{Fig_46tess_fractal_logical},   \\
    fractal-tree wedge,   \figref{Fig_46tess_x_topo}
	\end{tabular}  
    &   \begin{tabular}{@{}c@{}} 
	pruned fractal trees,   \figref{Fig_46tess_fractal_1fracton},   \\
     wedge corner, \figref{Fig_46tess_1fracton}
	\end{tabular} 
	\\  
	\bottomrule
	\end{tabular}
\end{table*}

\prlsection{Treeons in $(4,6)$  Y-cube   model}
Let us now discuss the excitations that result from $Z$ operators acting on the in-plane edges in the  $(4,6)$  Y-cube model.
%
A single in-plane $Z$ operator creates two composite
excitations,
each consisting of two excited vertex operators, as shown in \figref{Fig_46tess_lineon_pair}.
The two excited vertex operators (inset of \figref{Fig_46tess_lineon_pair})  share three in-plane edges.
Acting with a $Z$ operator on one of these shared edges will 
move the composite excitation to the adjacent vertex on the other side of the edge.
The excitation cannot move along other edges without creating additional  excitations.
%

Repeating this procedure,
we find that this composite excitation 
can move anywhere on the fractal tree shown in \figref{Fig_46tess_lineon_move}.
The construction of such a  tree is the same   as  $T_X$ in \figref{Fig_46tess_fractal_logical}. 
This composite excitation is similar to a lineon, except at each vertex it can choose between multiple fractal paths.
We call this new kind of mobility-restricted excitation a \emph{treeon} since its mobility is restricted to a fractal tree.


%

One non-trivial consequence of the \treeon s 
is how they form  logical operators.
%
A treeon  can travel from any one of the   many branches of the tree  to any other  one.
The product of the $Z$ operators along any such path
is then a  logical operator
(assuming rough boundary conditions that condense treeons).
One example is shown in \figref{Fig_46tess_lineon_logical}.
\\

\prlsection{Generalization to all tessellations}
Let us now summarize some properties of the hyperbolic X/Y-cube models on all $(p,q)$ tessellations with even $q$.
When $q\ge 6$, X-cube lineons  are replaced by a new type of excitations, treeons, that move on the fractal tree.
The $q=4$ case can be viewed as a special limit of a tree with only two branches at each vertex, which becomes a geodesic.
In this limit the \treeon s become lineons, and fracton dipoles gain mobility along the geodesics.
The properties of different hyperbolic lattices are summarized in Table.~\ref{TABLE_hyperbolic_X}.

\begin{figure}
	\centering
  \subfloat[\label{Fig_64tess_x_topo}]{\includegraphics[width=0.33 \columnwidth]{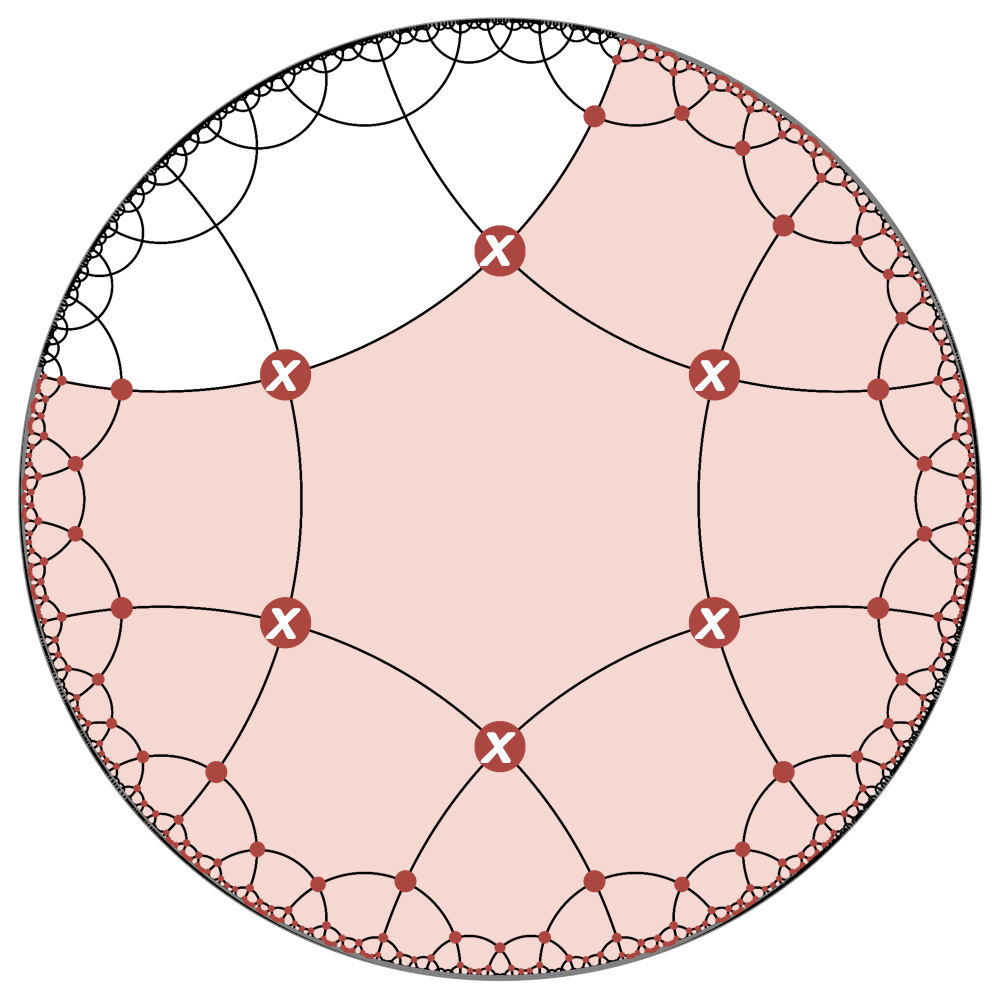}}\qquad
\subfloat[\label{Fig_64tess_fracton}]{\includegraphics[width=0.33 \columnwidth]{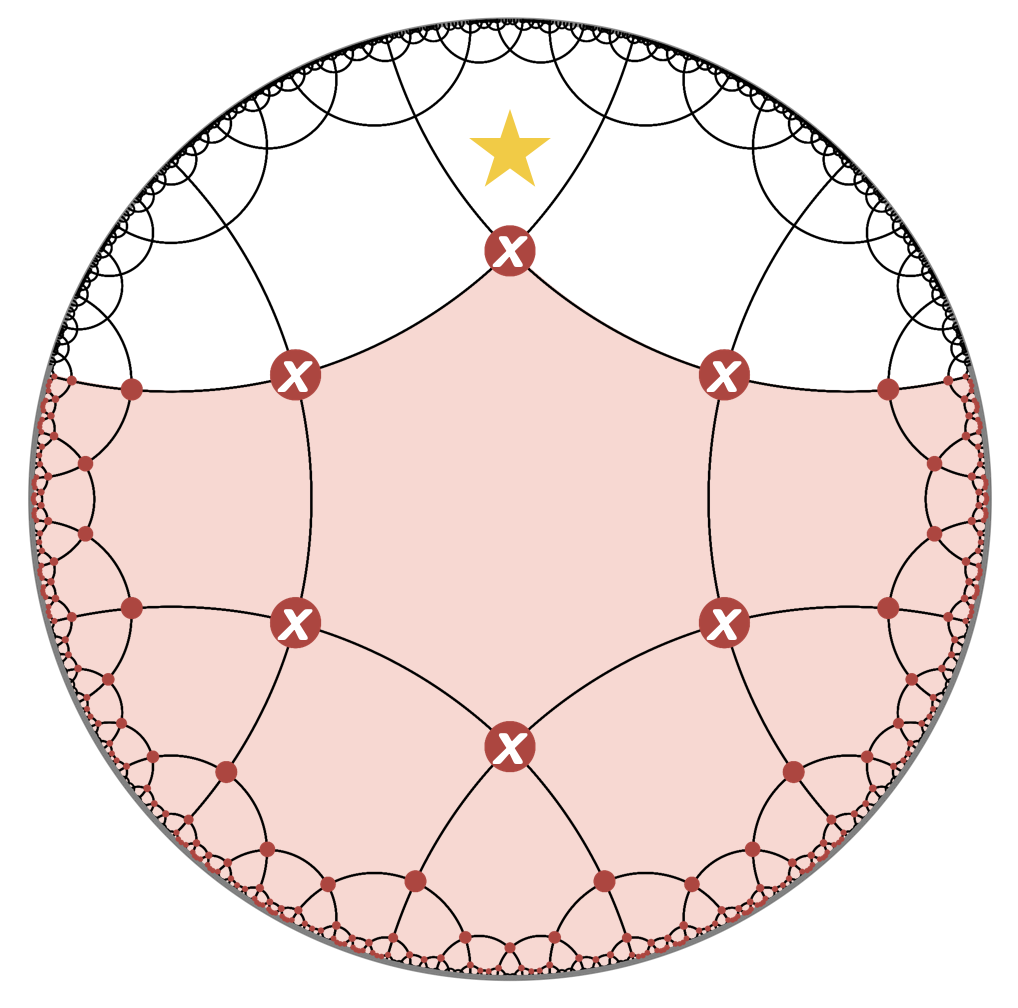}}	 
\caption{	\label{Fig_64tess_x}
(a) A logical membrane $X$ operator as the geodesic wedge for the $(6,4)$ model.
(b) A membrane $X$ operator with a corner that creates a fracton.
} 
\end{figure}

When $p$ is odd, a fracton can be created at the end of a series of truncated geodesics or fractal trees [Figs.~\ref{Fig_46tess_fractal_1fracton},\ref{Fig_54tess_1fracton}].
When $p$ is even,  logical $X$ membrane operators are allowed
in the shape of fractal-tree wedges for $q \geq 6$ [\figref{Fig_46tess_1fracton}] or geodesic wedges for $q = 4$ [\figref{Fig_64tess_x_topo}]. 
The intersection of two of these logical operators creates a single fractons at the corner [\figsref{Fig_46tess_1fracton} and \ref{Fig_64tess_fracton}].

Finally, tessellations of $1/p + 1/q =1/2$ are special limits of the embedding space becoming  flat rather than hyperbolic.
In the case of $p=q=4$ (square lattice)  we recover the 3D X-cube model on a cubic lattice.
When $(p = 3,q = 6)$, the 2D tessellation forms a triangular lattice,
and we can define the Y-cube model on a stack of these triangular lattices.
In this case, the  Y-cube model treeons  we encountered for $(p>3, q= 6)$ become planeons that move on a honeycomb network embedded in the flat triangular layer.
This results because when the hyperbolic geometry is made flat,
the fractal tree that the treeon can traverse collapses onto itself and reduces to a 2D honeycomb,
see the
\supp
for more detail (although we leave the in-depth study of this triangular model to future work). \\

\prlsection{Summary and outlook}
We introduced the Y-cube model on stacked tessellations of the hyperbolic plane.
We discovered that the Y-cube model features a new kind of particle with restricted mobility: a treeon, which is constrained to move along a fractal tree.
We also find that in the hyperbolic X-cube and Y-cube models with even $p$, fractons can be created by either in-plane membrane or fractal operators (\figsref{Fig_46tess_fractal_1fracton} and \ref{Fig_46tess_x}).
We are not aware of any previously studied models with this property.

These models also serve as a concrete  example of how the lattice geometry (tessellation) of fracton orders can
determine their fundamental properties, even when the embedding space (in this case $H_2\times S^1$) is the same.
Our discovery suggests that there are still many  fracton orders with new and exotic features not seen before,
especially when their underlying graphs/lattices are beyond the flat space ones, i.e., lattices with  
 no translational symmetry.
Like Type-II fracton orders, the hyperbolic Y-cube models are not foliated fracton orders \cite{ShirleyFoliation,fractonEntanglement}, 
challenging us for new insight of classification schemes of fracton order \cite{BifurcatingEntanglement}.


This work provides one of the simplest examples of fracton order beyond the flat space,
but there is much room left for future exploration.
One future topic is to impose boundary conditions on the hyperbolic plane and study the ground state degeneracy, logical operators, and quantum information encoding.
%
It is also useful to ask if the new physics from the hyperbolic structure provides  benefits in quantum memory storage.
Another  direction is to investigate fracton models on the 3D hyperbolic space $H_3$, or general graphs without translational symmetry \cite{ShuHengComplexityGraphs}.\\

H.Y. and A.H.N. were supported by the National Science Foundation Division of Materials Research under the Award DMR-1917511.
K.S. was partially supported by
  the Walter Burke Institute for Theoretical Physics at Caltech; and
  the U.S. Department of Energy, Office of Science, National Quantum Information Science Research Centers, Quantum Science Center.\\

\bibliography{fractons,title_date_Han_Yan}


\clearpage
\setcounter{equation}{0}
\setcounter{figure}{0}
\setcounter{table}{0}
\makeatletter
\renewcommand{\theequation}{S\arabic{equation}}
\renewcommand{\thefigure}{S\arabic{figure}}
\renewcommand{\bibnumfmt}[1]{[#1]}
\renewcommand{\citenumfont}[1]{#1}


\begin{widetext}
\begin{center}
	\Large{\textbf{Supplementary Materials for ``\thistitle"}}
\end{center}
\end{widetext}


\section{The case of $(3,6)$  tessellation }
\label{app:triangular}


Here, we discuss some properties of the Y-cube model on the $(3,6)$ tessellation (triangular lattice) $\times S^1$.
This is a special limit of the $(p,q)$ tessellations that is geometrically flat instead of  hyperbolic,
  which drastically affects the mobility properties of the excitations.
We leave a more complete study of this model to future work.
  
\begin{figure}[bh!]
	\centering 
 \includegraphics[width=0.48\textwidth]{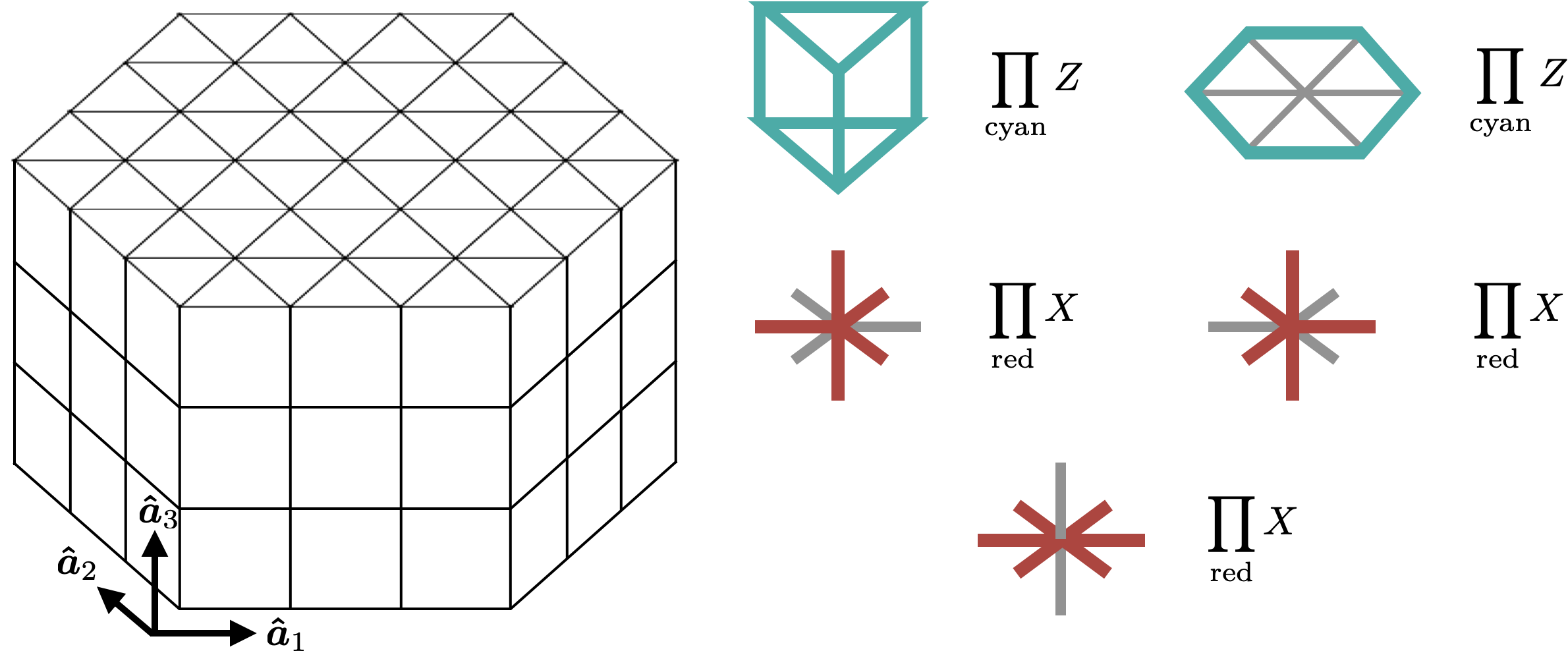}
\caption{\label{Fig_triangle_Hamiltonian}
The triangular$\times S^1$ lattice and its Hamiltonian terms.
} 
\end{figure}

The Hamiltonian consists of three types of terms,
  shown in \figref{Fig_triangle_Hamiltonian}.
The first two types are the product of $Z$'s on the edges of a triangle prism, and certain products of $X$'s around   vertices.
On this flat tessellation, an additional term can be added to the Hamiltonian,
  which is not analogous to any term in hyperbolic tessellations: namely 
the product of $Z$'s around a hexagon.
The existence of this third term only on flat space is related to the fact (explained below) that $Z$ operators on in-plane edges create planons,
  rather than treeons as in the hyperbolic Y-cube model.

\begin{figure}
	\centering 
 \subfloat[\label{Fig_triangle_treeon_pair}]{\includegraphics[width=0.39\columnwidth]{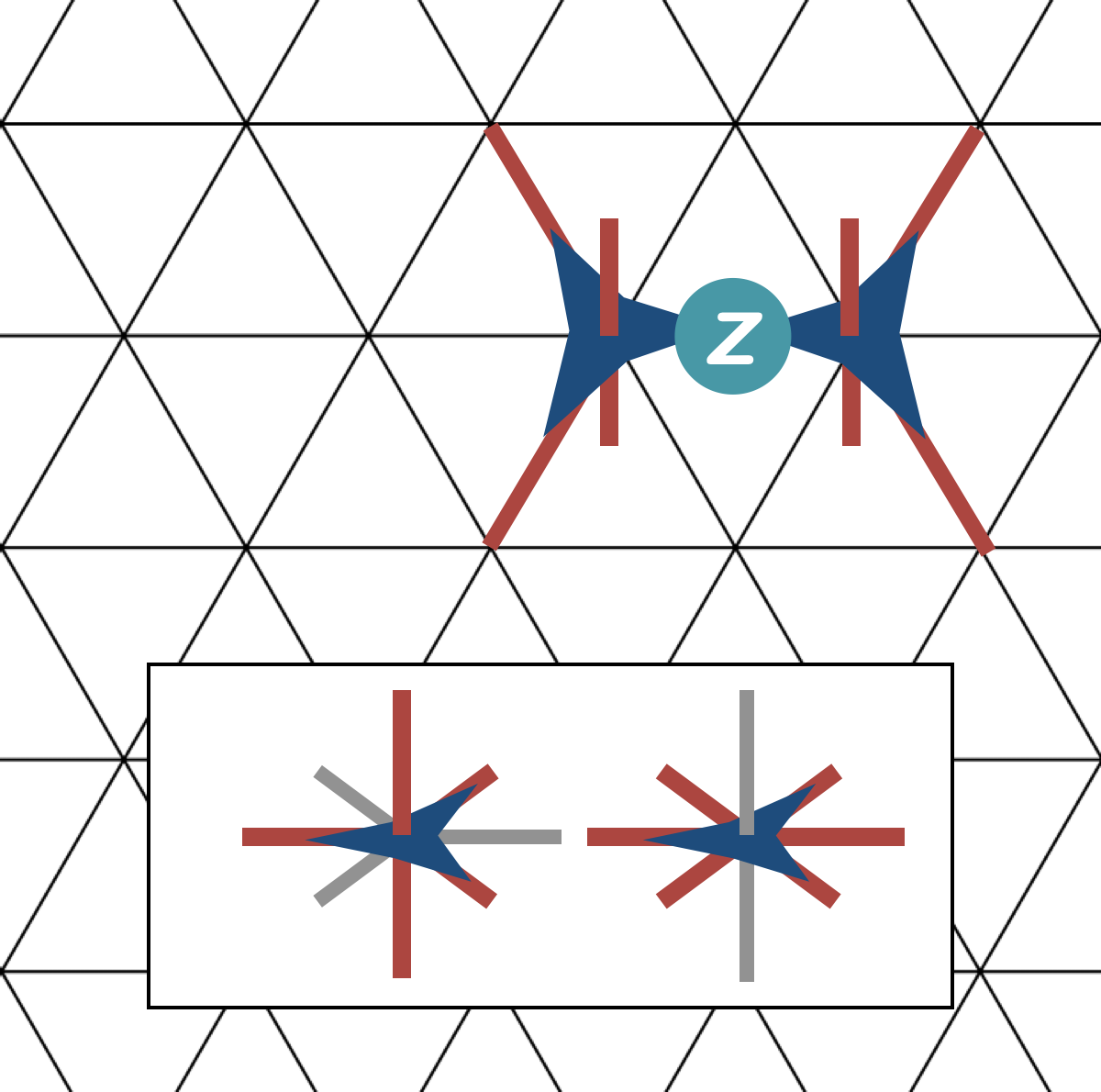}}
 \qquad
 \subfloat[\label{Fig_triangle_treeon_path}]{\includegraphics[width=0.39  \columnwidth]{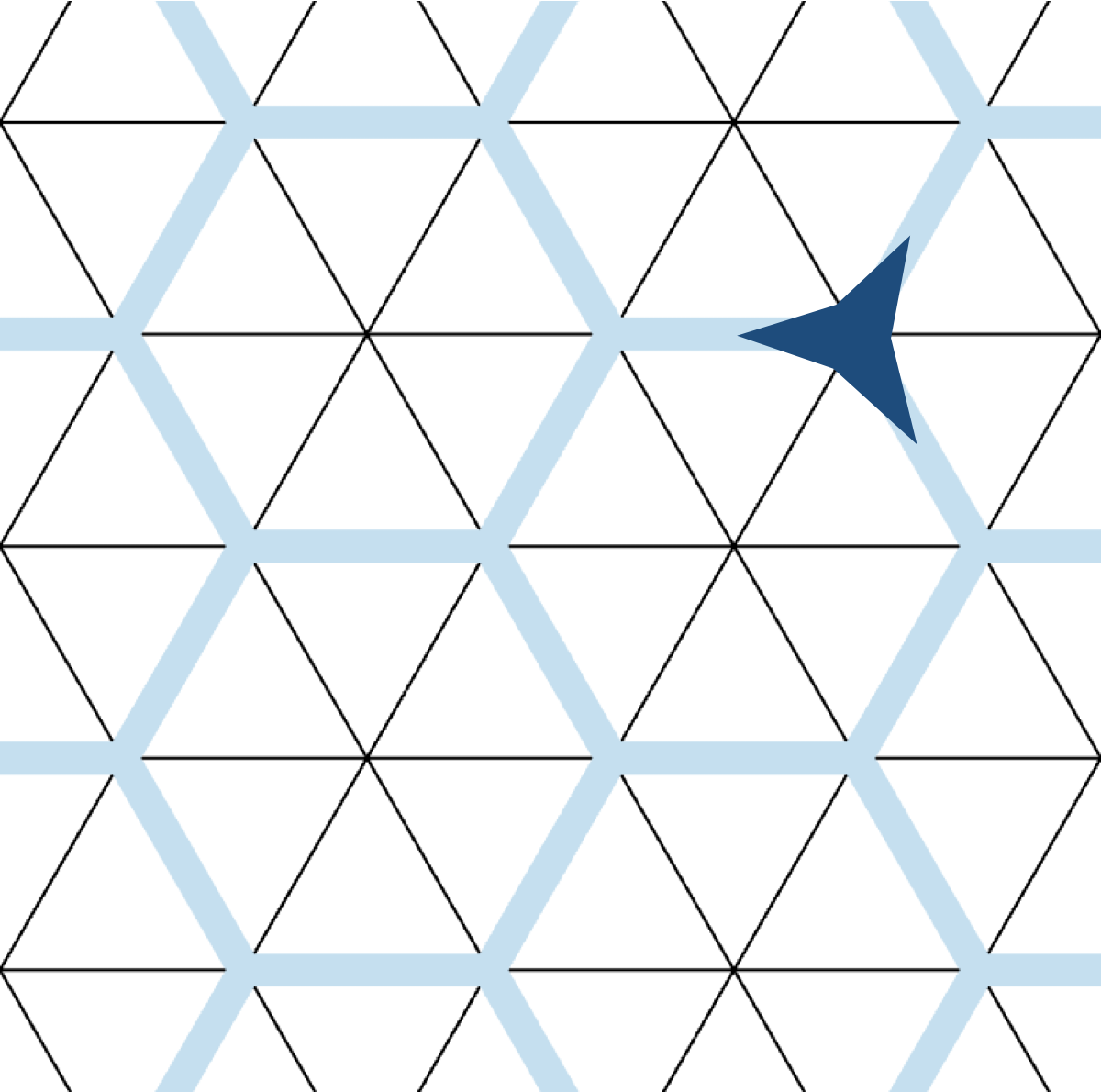}}
\caption{\label{Fig_triangle}
(a) A $Z$ operator (teal) acting on an in-plane edge creates two planeons (blue stars).
The inset shows the two excited terms in the Hamiltonian for a single planeon.
(b) A single planeon can travel on the hexagonal sublattice of the triangular lattice, colored in light blue.
} 
\end{figure}

We first examine the excited states of the vertex terms. 
Recall from the main text that when $p > 3$ and $q = 6$,
  $Z$ operators acting on in-plane edges create treeon excitations that are restricted to move on a fractal tree. 
However on the $(3,6)$ tessellation,
  we find that these vertex excitations are instead planeons. This can be seen as follows.
Locally, a $Z$ operator on an in-plane edge creates two such planeons [\figref{Fig_triangle_treeon_pair}]. 
Each planeon is in the excited state of two vertex operators [inset of \figref{Fig_triangle_treeon_pair}].
Similar to a treeon, these planeons can move along any one of three in-plane edges connected to its vertex.
However since the lattice is no longer hyperbolic,
  this mobility results in the mobility of a planeon,
  which can move within a hexgonal sublattice, as shown in \figref{Fig_triangle_treeon_path}.
There are three flavors of this planeon,
  one flavor for each hexagonal sublattice.
The flavor of the planeon can be changed at the expense of creating the fully mobile excitation described in the following paragraph.

\begin{figure}
	\centering 
 \subfloat[\label{Fig_triangle_free1}]{\includegraphics[width=0.35\columnwidth]{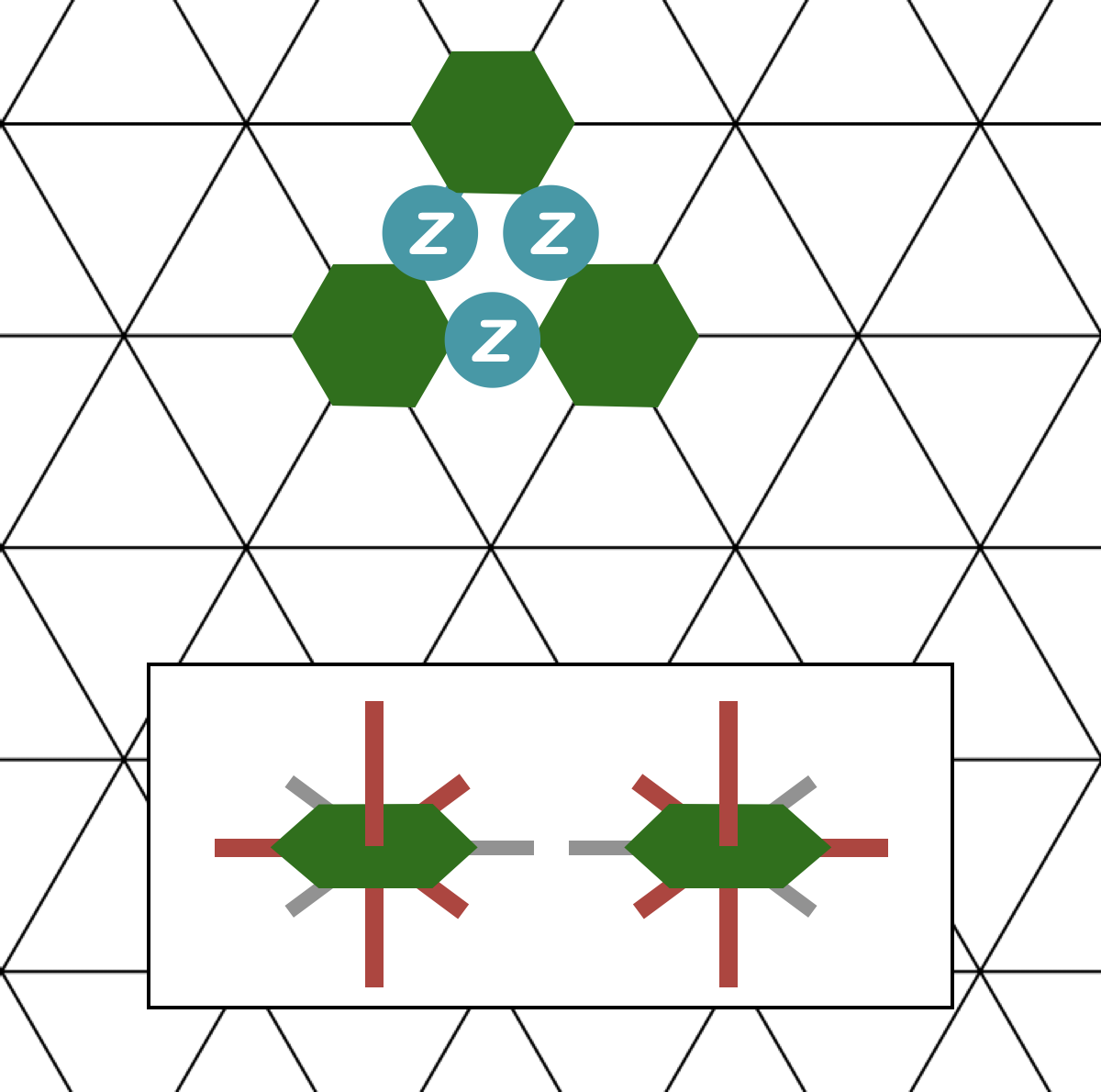}}
 \qquad
 \subfloat[\label{Fig_triangle_free2}]{\includegraphics[width=0.35  \columnwidth]{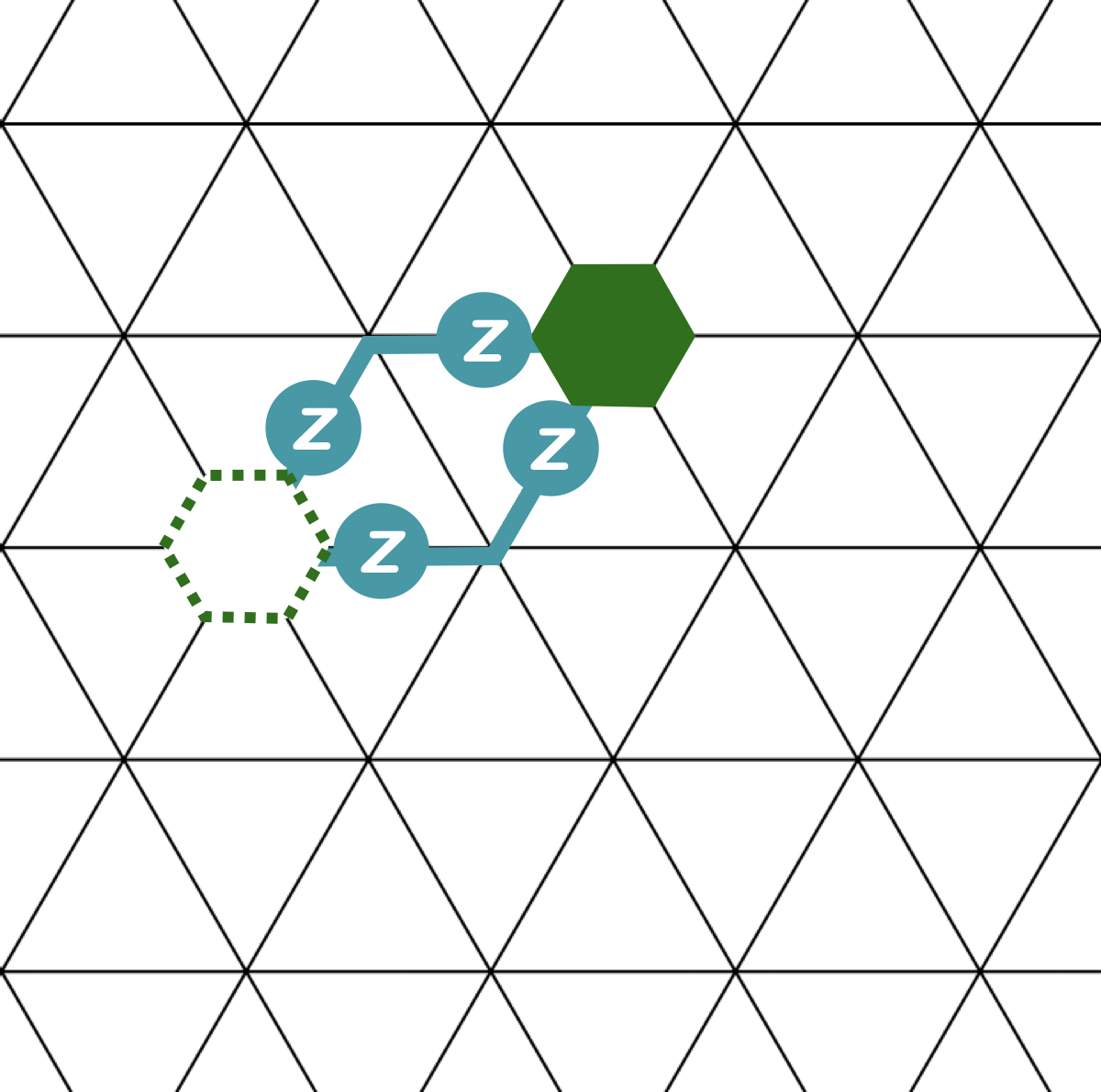}} \qquad
 \subfloat[\label{Fig_triangle_free3}]{\includegraphics[width=0.45  \columnwidth]{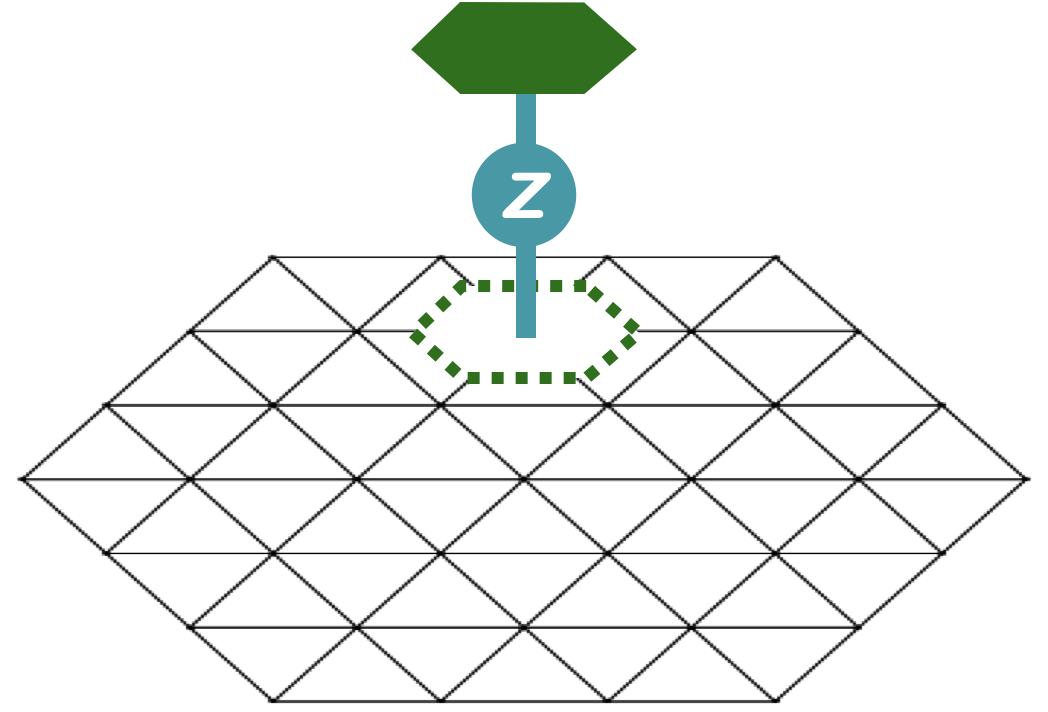}}
\caption{\label{Fig_triangle_free}
(a) Three $Z$ operators (teal) acting on an in-plane triangle creates three excitations (green hexagons).
The inset shows the two excited terms in the Hamiltonian for a single excitation.
(b) A single excitation can travel on a 
sub-triangular lattice of the original triangular lattice.
This movement is done using four $Z$ operators shown on the figure.
(c) A single excitation can also travel vertically.
This movement is done using a $Z$ operator on the vertical link shown on the figure.
} 
\end{figure}

In hyperbolic geometry, 
the composite excited state of two vertex operators created by an out-of-plane $Z$ operator is a lineon that can move in the vertical direction only. 
On the $(3,6)$ tessellation,
  the analogous excitation is instead free to move in 3D. 
Three $Z$ operators on the edges of an in-plane triangle creates three  such excitations [\figref{Fig_triangle_free1}].
Each composite excitation is in the excited state of two vertex operators [inset of \figref{Fig_triangle_free1}].
Unlike on the hyperbolic plane, the flat geometry allows these
 composite excitations to move along six in-plane directions, as shown in \figref{Fig_triangle_free2}.
Repeated in-plane movement can span a triangular sub-lattice on the original lattice.
These composite excitations can also move along the vertical direction as shown in \figref{Fig_triangle_free3}, giving them 3D mobility.

\begin{figure}
	\centering 
 \includegraphics[width=0.35\textwidth]{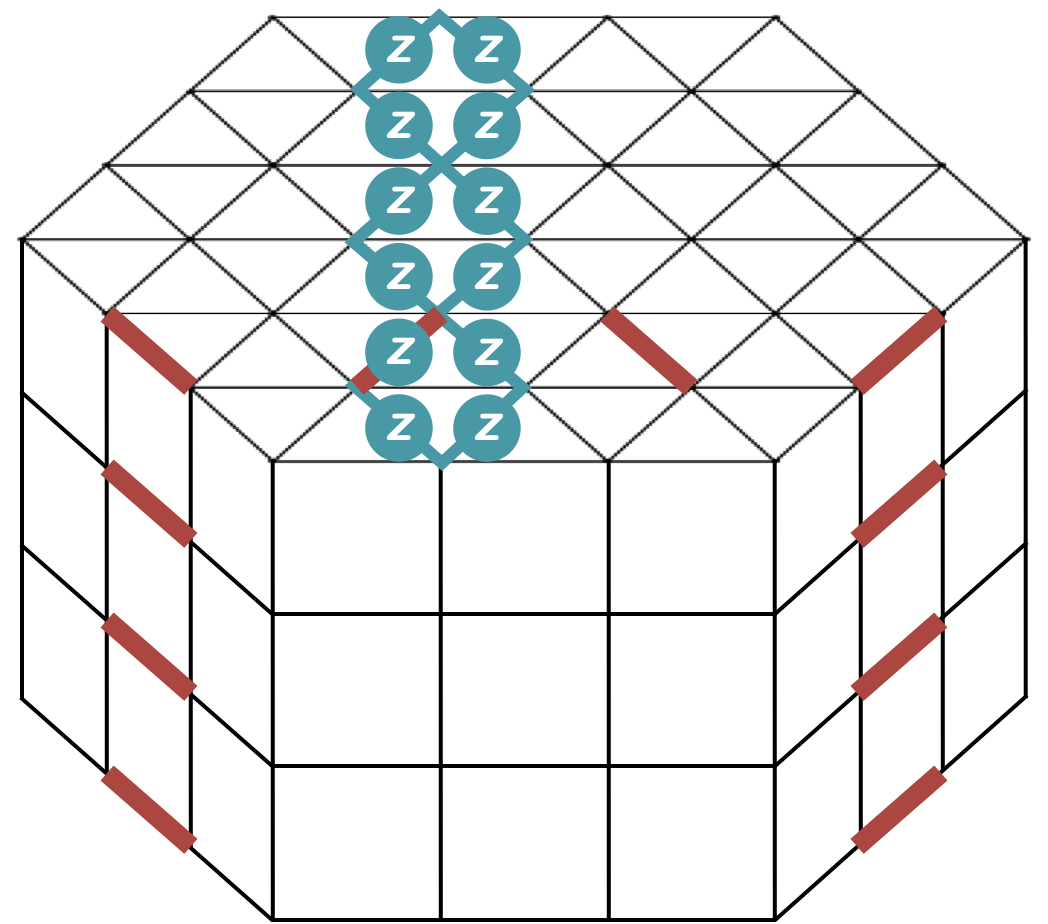}
\caption{\label{Fig_triangle_free_membrane}
A flux operator consisting of a product of $X$ operators (red) within a 2D membrane.
The flux operator anticommutes with the string operator (\figref{Fig_triangle_free2}) of the mobile charge excitation.
} 
\end{figure}

These composite vertex excitations are similar to 3D toric code charges. Naively, there are three flavors: one on each of the 3 sublattices. 
But by acting with a triangle of $Z$ operators [\figref{Fig_triangle_free1}],  the composite of three flavors annihilate. 
Thus, there are actually only two independent flavors.
The corresponding   flux operator (similar to a 3D toric code flux operator)
is the membrane operator consisting of an out-of-plane stack of products of $X$ operators on the red links in \figref{Fig_triangle_free_membrane}. 
This 2D membrane operator creates a loop excitation around its boundary.  
Presumably, there are two flavors of this membrane operator.

\begin{figure}[b!]
	\centering 
 \includegraphics[width=0.4\textwidth]{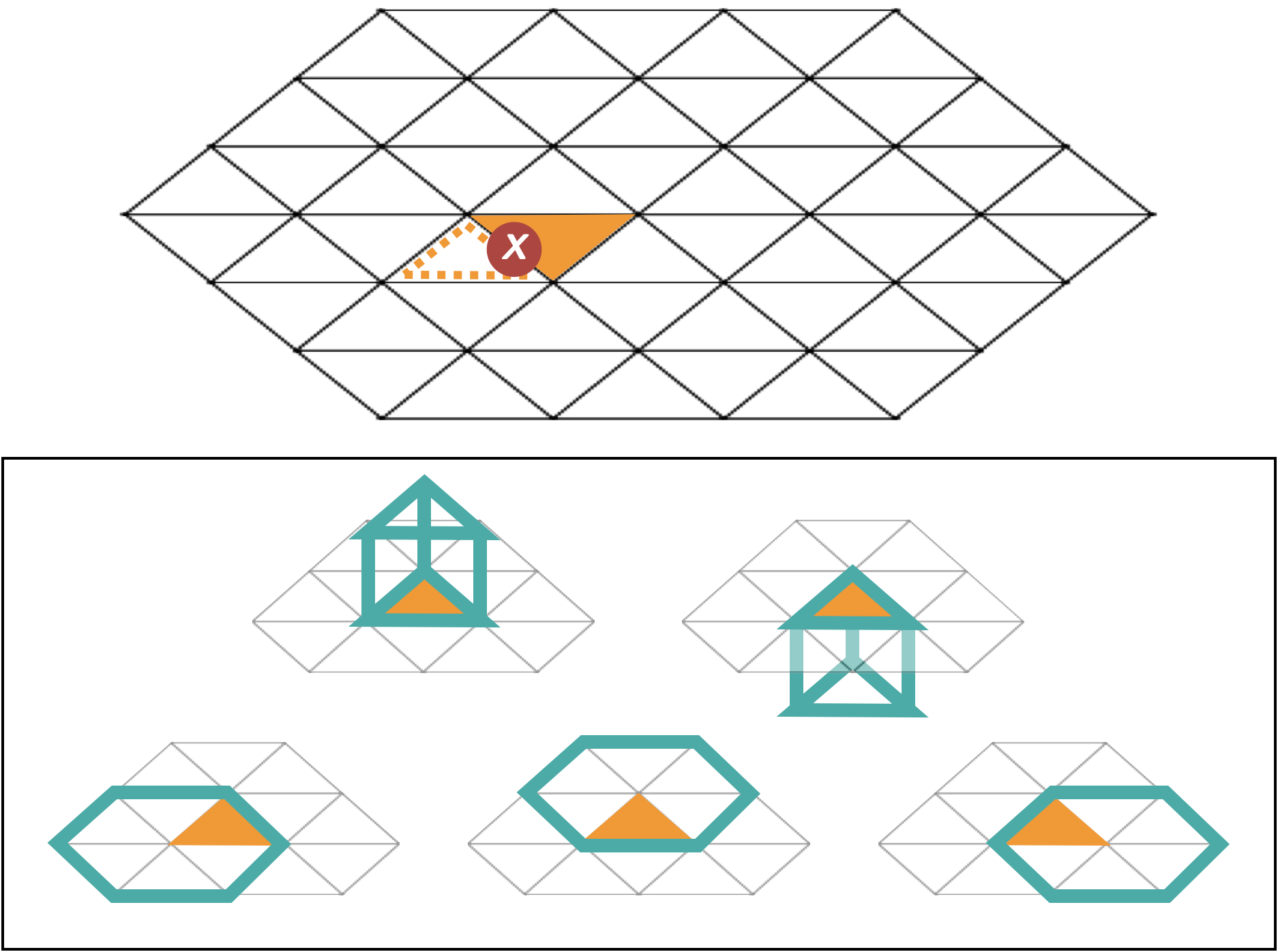}
\caption{\label{Fig_triangle_prismplaneon}
The action of an in-plane $X$ operator moves a planeon (orange triangle in the top panel).
The planeon is a composite excitations of the five $Z$ terms (teal) shown in the lower panel.
} 
\end{figure}

The action of in-plane $X$ operators creates planeon excitations with in-plane mobility.
The planeon is a composite excitations of the $Z$ terms shown in \figref{Fig_triangle_prismplaneon}.
This planeon anticommutes with the planeon shown in \figref{Fig_triangle}.

Finally, we note that for an $L \times L \times L$ periodic lattice (with lattice constants $\bm{\hat{a}}_{1,2,3}$ shown in  \figref{Fig_triangle_Hamiltonian}),
  the ground state degeneracy is GSD = $2^{6+2L}$
  when $L$ is multiple  of $3$ (for which different flavors of particles do not turn into each other due to boundary conditions).

Therefore, this model supports a pair of anticommuting planeons on each layer, similar to stacks of toric code.
The model also supports two sets of fully mobile charges along with flux-line excitations, similar to two copies of 3D toric code.
Thus, it seems plausible that the model is local-unitary equivalent \cite{XieGLU} to the following hybrid fracton order \cite{hybridFracton}: two copies of 3D toric code and decoupled stacks of 2D toric codes.
Indeed, the ground state degeneracy is also consistent with this possibility.
We leave the resolution of this possibility to future work.

\clearpage

\end{document}